\documentclass[manuscript]{aastex}

\usepackage{natbib}
\usepackage{longtable}
\usepackage{epsfig}
\usepackage{amssymb}
\usepackage{amsmath}

\newcommand{\ep}{\epsilon_{\rm p}}
\newcommand{\es}{\epsilon_{\rm s}}
\newcommand{\be}{\begin{equation}}
\newcommand{\ee}{\end{equation}}

\slugcomment{submitted to ApJ}

\shorttitle{BORG spec}
\shortauthors{Treu et al.\ }

\begin{document}

\title{Inferences on the distribution of Lyman $\alpha$ emission of $z\sim7$ and $z\sim8$ galaxies}

\author{Tommaso Treu\altaffilmark{1,2}, Michele Trenti\altaffilmark{3}, Massimo Stiavelli\altaffilmark{4}, Matthew W. Auger\altaffilmark{1}, Larry D.~Bradley\altaffilmark{4}}

\altaffiltext{1}{Department of Physics, University of California,
Santa Barbara, CA 93106, USA ({\tt tt@physics.ucsb.edu})}
\altaffiltext{2}{Packard Fellow}
\altaffiltext{3}{University of Colorado, Center for Astrophysics and Space Astronomy, 389-UCB, Boulder, CO 80309, USA}
\altaffiltext{4}{Space Telescope Science Institute, 3700 San Martin Dr, Baltimore MD, 21218}

\begin{abstract}
Spectroscopic confirmation of galaxies at $z\sim7$ and above has been
extremely difficult, owing to a drop in intensity of Lyman$\alpha$
emission in comparison with samples at $z\sim6$. This crucial finding
could potentially signal the ending of cosmic reionization. However it
is based on small datasets, often incomplete and heterogeneous in
nature. We introduce a flexible Bayesian framework, useful to
interpret such evidence. Within this framework, we implement two
simple phenomenological models: a smooth one, where the distribution
of Lyman$\alpha$ is attenuated by a factor $\es$ with respect to
$z\sim6$; a patchy one where a fraction $\ep$ is absorbed/non-emitted
while the rest is unabsorbed. From a compilation of 39 observed
$z\sim7$ galaxies we find $\es=0.69\pm0.12$ and $\ep=0.66\pm0.16$. The
models can be used to compute fractions of emitters above any
equivalent width $W$. For $W>25$\AA, we find
$X^{25}_{z=7}=0.37\pm0.11$ ($0.14\pm0.06$) for galaxies fainter
(brighter) than M$_{\rm UV}$=-20.25 for the patchy model, consistent
with previous work, but with smaller uncertainties by virtue of our
full use of the data. At $z\sim8$ we combine new deep (5$\sigma$ flux
limit $10^{-17}$ergs$^{-1}$cm$^{-2}$) Keck-NIRSPEC observations of a
bright $Y$-dropout identified by our BoRG Survey, with those of three
objects from the literature and find that the inference is
inconclusive. We compute predictions for future near-infrared
spectroscopic surveys and show that it is challenging but feasible to
constrain the distribution of Lyman$\alpha$ emitters at $z\sim8$ and
distinguish between models.
\end{abstract}

\keywords{gravitational lensing --- galaxies: evolution --- galaxies: high-redshift}

\section{Introduction}

One of the frontiers of modern cosmology is cosmic reionization. When
did it occur? What sources of light provided enough UV photons to
reionize the universe and end cosmic dark ages? Clues like the cosmic
microwave background \citep{Kom++11}, the luminosity function of
high-z quasars and the Gunn-Peterson effect \citep{Fan++06b}, suggest
that reionization occurred between $z\sim8$ and $z\sim12$, caused by
the UV emission of the first galaxies \citep[see, e.g.,][for recent
reviews]{Sti09,Rob++10}.

The commissioning of the Wide Field Camera 3 on board the Hubble Space
Telescope -- with orders of magnitude more discovery potential than
the previous infrared camera NICMOS -- and of high sensitivity wide
field near infrared imagers like HAWK-I from the ground, has opened up
the wholesale study of the universe beyond $z\sim7$, when
Lyman-$\alpha$ is completely redshifted into the near infrared.

The first studies based on the dropout technique \citep{Ste++96} to
identify galaxies at $z\sim7,8$ and beyond, are consistent with a UV
luminosity function with a steep faint end (slope close to $-2$) and a
characteristic magnitude significantly fainter than at lower
redshifts~\citep[e.g.,][]{Bou++11b,Cas++10a,Cas++10b}. This indicates
that the overall luminosity and star formation rate of galaxies at
$z\sim8$ is much lower than at $z\lesssim6$ and dominated by the
fainter galaxies. The jury is still out on whether the sources are
enough to reionize the universe \citep[e.g.][]{Lor++11,Trenti++10}.

A key issue however, is that of spectroscopic follow-up. This is
essential for two reasons. On the one hand, spectroscopic confirmation
of at least a subset of the sources is needed to prove beyond any
reasonable doubt that dropout selected galaxies are indeed at high
redshift, and verify the low contamination rates suggested by
simulations of imaging searches. On the other hand, spectroscopic
information on the intensity and shape of Lyman$\alpha$ emission
constrains the properties of star formation in early galaxies and the
radiative transfer properties of their interstellar medium and
surrounding intergalactic medium, which in turn provides information
on the geometry and physics of reionization.

Significant progress has been made to date, especially at $z\sim7$,
where high-sensitivity multiplexed optical spectrographs can be used
to reach sensitivity to Lyman$\alpha$ equivalent widths of only a few
\AA\ for several sources at a time. Recent studies of $z\sim7$
galaxies report a very interesting result, which might provide a vital
clue for reconstructing the history of cosmic reionization. Whereas
the fraction of dropouts that are Lyman $\alpha$ emitters, increases
steadily out to $z\sim 6$ \citep{SEO11,Cur++11}, at $z\sim7$ the
fraction appears to decline significantly
\citep{Fon++10,Sch++11,Ono++11,Pen++11}, possibly signaling a change
in the opacity of the intergalactic medium.  Narrow band searches for
Lyman $\alpha$ emitters provide a consistent picture
\citep{Kas++06,Ouc++10,Hu++10,Cle++11}.

Beyond $z\sim8$ spectroscopic follow-up has been much more limited,
owing to the challenges of observing in the near infrared and the lack
of multiplexing capabilities of current generations of infrared
spectrographs at Keck and VLT. A few detections have been reported in
the literature~\citep[e.g.,][]{Leh++10,Sta++07}, but they are of
marginal significance and lack independent confirmation (Bunker et
al.\ 2011, in preparation). Part of the difficulty of following up
$z\sim8$ galaxies also arises from the fact that most of the
candidates so far have been identified from deep and rather narrow
WFC3 searches, resulting in very faint sources that can be confirmed
from the ground only for extraordinarily high Lyman$\alpha$ equivalent
widths.

Identifying and following up relatively bright $z\sim8$ Y-dropouts is
the main goal of the Brightest of Reionization Galaxies Survey
\citep[hereafter BoRG][]{BORG1}. By means of pure parallel
observations (GO-11700 and 12572; PI Trenti), the BoRG survey is
collecting hundreds of square arcminutes of WFC3 images optimized for
$z\sim8$ galaxies detection, completing nicely searches in legacy
fields like CANDELS (PIs: Faber \& Ferguson). The first results
include the detection of four bright candidates \citep{BORG1} as well
as an overdensity of fainter dropouts in one of the
fields~\citep{BORG2}. Another newly discovered bright candidate is
presented in this paper.

Further progress in identifying new bright candidates is expected from
systematic deep surveys of the legacy fields as well as imaging the
fields of clusters of galaxies exploiting lensing magnification. With
new multiplexed infrared spectrographs like MOSFIRE \citep{McL++10}
expected to be commissioned soon, it is reasonable to assume that the
flux of spectroscopic data will increase significantly in the next few
years. However, as the observations are challenging and require
considerable investment, it is also likely that the information that
will be acquired and published will be heterogeneous in depth,
wavelength coverage, significance, and sample selection.

This paper is concerned with introducing a simple yet powerful
Bayesian formalism that allows one to combine in an efficient and
rigorous manner spectroscopic data heterogeneous in nature to infer
the distribution of Lyman $\alpha$ intensity at high redshift. The
formalism is able to deal with spectra with noise varying as a
function of wavelength, with incomplete wavelength coverage
incorporating the information from photometric redshifts, with
detections and non-detections. For any set of model of the intrinsic
distribution of Lyman-$\alpha$ equivalent width at a given redshift,
the method provides posterior probability distribution functions for
the model parameters as well as the evidence that can be used to
perform model selection.

We illustrate this framework by implementing two simple models of
Lyman $\alpha$ distribution, based on that observed at $z\sim6$ and
meant to represent two simple idealized scenarios of reionization. In
the first model, dubbed ``patchy'' absorption, the distribution of
Lyman$\alpha$ intensity is the same as at $z\sim6$ for $\ep$ sources,
while the others are either completely absorbed or do not emit. In the
second model, dubbed ``smooth'' absorption, Lyman$\alpha$ is quenched
for all line of sights by a factor $\es$. The parameters $\ep$ and
$\es$ can be physically interpreted as the average excess optical
depth of Lyman$\alpha$ with respect to $z\sim6$, i.e. $\langle
e^{-\tau_{Ly\alpha}}\rangle$.

Even though clearly these toy models do not include the physics that
is used to compute real models \citep{DMW11,D+F11}, they should
somewhat bracket reality, where we expect a distribution of absorption
along different lines of sight, and overall a non-zero smooth
component.  The patchy model represents a zero-th order idealization
of the complex topology of the reionization process inferred from
cosmological simulations, so that galaxies at the same redshift can be
surrounded by IGM with different ionization state depending on their
environment and past star formation history
\citep{Ili++06,McQ++07,Shi++08}. In this approach, the patchiness of
the absorption is also likely to depend on the luminosity and rarity
of the sources \citep[e.g.][]{Fur++06}. In reality, even in patchy
reionization, the distribution of lyman $\alpha$ optical depths will
be closer to a gaussian, and certainly not bimodal as in our
simplified model \cite[][and references therein]{DMW11}. In this sense
our model represents and extreme idealization of patchy
reionization. The smooth absorption model represents instead a simpler
approach often adopted in analytical models of reionization, where the
evolution of the ionized fraction in the Universe is assumed to be
spatially uniform on average and linked to the observed number of
ionizing photons \citep[e.g.,][]{Sti++04,B+H07,S+V08,Trenti++10}.  In
reality, smooth reionization models will clearly not be characterized
by a delta function in optical depth, but a distribution with smaller
variance then the one appropriate for patchy models. Thus our smooth
reionization model is an extreme idealization with zero
variance. Thus, in this sense our two models taken together bracket
the range expected for realistic physical models.

We then apply these models to data at $z\sim7$ and $z\sim8$. At
$z\sim7$ we analyze a sample of 39 deep observations from
the~literature~\citep{Pen++11,Ono++11,Sch++11}. At $z\sim8$ we apply
our methodology to new deep Keck observations of a bright dropout
identified by the BoRG Survey, as well to a sample of 3 additional
objects taken from the literature for which deep infrared observations
are available: two objects from the paper by~\citet[][including a
target from BoRG]{Sch++11} and the detection reported by
\citet{Leh++10}. The model is then used to compute forecasts, useful
for planning future near-infrared observing campaigns.
 
The paper is organized as follows. In~\S~\ref{sec:method} we describe
our method. In~\S~\ref{sec:obs} we present new
observations. In~\S~\ref{sec:results} we derive current limits on the
distribution of Lyman$\alpha$ at $z\sim7$ and $8$ and compare with
previous work. In~\S~\ref{sec:predictions} we present our
forecasts. Section~\ref{sec:conc} concludes and summarizes the paper.

We assume a concordance cosmology with matter and dark energy density
$\Omega_m=0.3$, $\Omega_{\Lambda}=0.7$, and Hubble constant
H$_0$=100$h$kms$^{-1}$Mpc$^{-1}$, with $h=0.7$ when necessary. Base-10
logarithms, AB magnitudes, and the cgs system are used unless
otherwise stated. For conciseness, we adopt the following shorthand
filter names $z'$ (ACS F850LP), $Y$ (WFC3-IR F098M), $J$ (WFC3-IR
F125W), $H$ (WFC3-IR F160W).


\section{Bayesian Inference}
\label{sec:method}

We now describe a general method that can be used to constrain the
distribution of equivalent width of Lyman $\alpha$, exploiting all the
information available, including non-detections, wavelength dependent
sensitivities, incomplete wavelength coverage, and photometric
redshift.

For the sake of simplicity we shall assume that the intrinsic
rest-frame equivalent width distribution is obtained by rescaling the
one measured at $z\sim 6$ by \citet{SEO11} $p_6(W)$. Note that this is
implicitly assumed by most studies of this topic
\citep{Sch++11,Fon++10,Pen++11,Ono++11}, 
and it is a very reasonable approach considering the dearth of
information.

As a practical matter, we describe the \citet{SEO11} distribution as
the sum of a truncated Gaussian plus a delta function. Given the
observational uncertainties, the Gaussian choice is by no means
unique, but it is sufficient for our purposes and computationally
convenient:

\begin{equation}
p_6(W)=\frac{2 A}{\sqrt{2 \pi}W_{c}}e^{-\frac{1}{2}\left(\frac{W}{W_{c}}\right)^2}H(W)+(1-A)\delta(W),
\end{equation}
with W$_{c}$=47\AA, A=0.38 for the brighter sources (-21.75$<$M$_{\rm
UV}<$-20.25) and W$_{c}$=47\AA, A=0.89 for the fainter sources
(-20.25$<$M$_{\rm UV}<$-18.75). A is the fraction of emitters and H is
the Heaviside step function.  Note that the term $(1-A)$ includes the
fraction of interlopers in dropout-selected samples. If the fraction
of interlopers changes with redshift, this can be easily be accounted
for in the evolutionary model, with a simple generalization (in the
patchy model, this is already accounted for in $\ep$).  Many
alternative parameterizations are possible. An alternative
parameterization of the $z\sim6$ distribution, similar to that adopted
by \citet{Pen++11} is described in the appendix, showing that the
specific choice of the parameterization contributes little to the
overall uncertainties at this point. Another possible parameterization
is the exponential adopted by \citet{D+W11}. The method is very
general and any parameterization of the $z\sim6$ distribution can be
implemented. As the samples at $z\sim6$ improve in size beyond the 74
galaxies in the \citet{SEO11} sample, it will be possible to restrict
the range of possible parametrizations and reduce the related
uncertainties.

\begin{figure}
\begin{center}
\resizebox{\columnwidth}{!}{\includegraphics{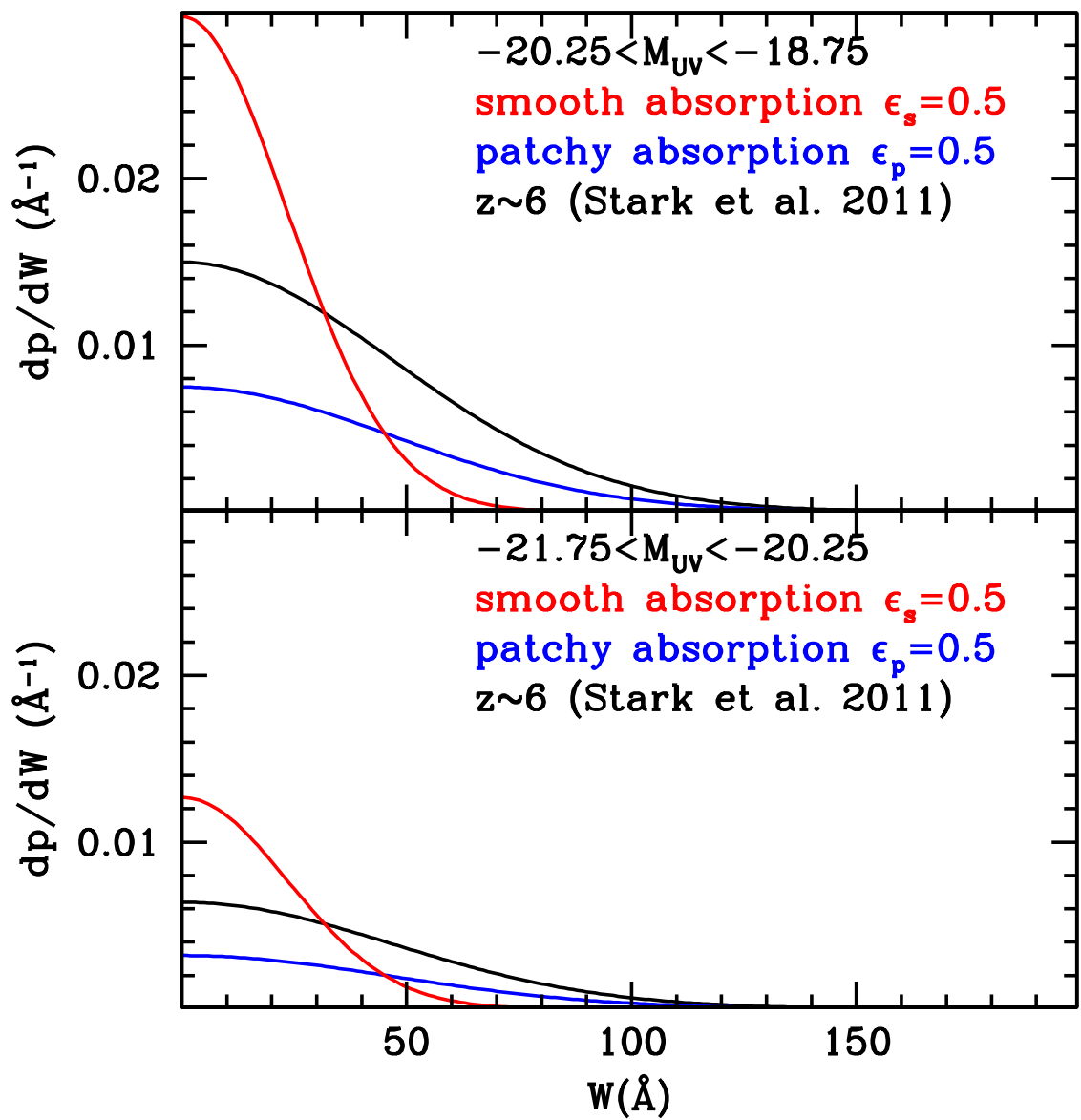}}
\end{center}
\figcaption{Illustration of our model intrinsic distribution of rest-frame equivalent width $W$. A fit to the distribution measured at $z\sim6$ by Stark et al. (2011) is shown as a black line. Red and blue lines represent a model with smooth and patchy absorption, respectively.
\label{fig:models}}
\end{figure}

\begin{figure}
\begin{center}
\resizebox{\columnwidth}{!}{\includegraphics{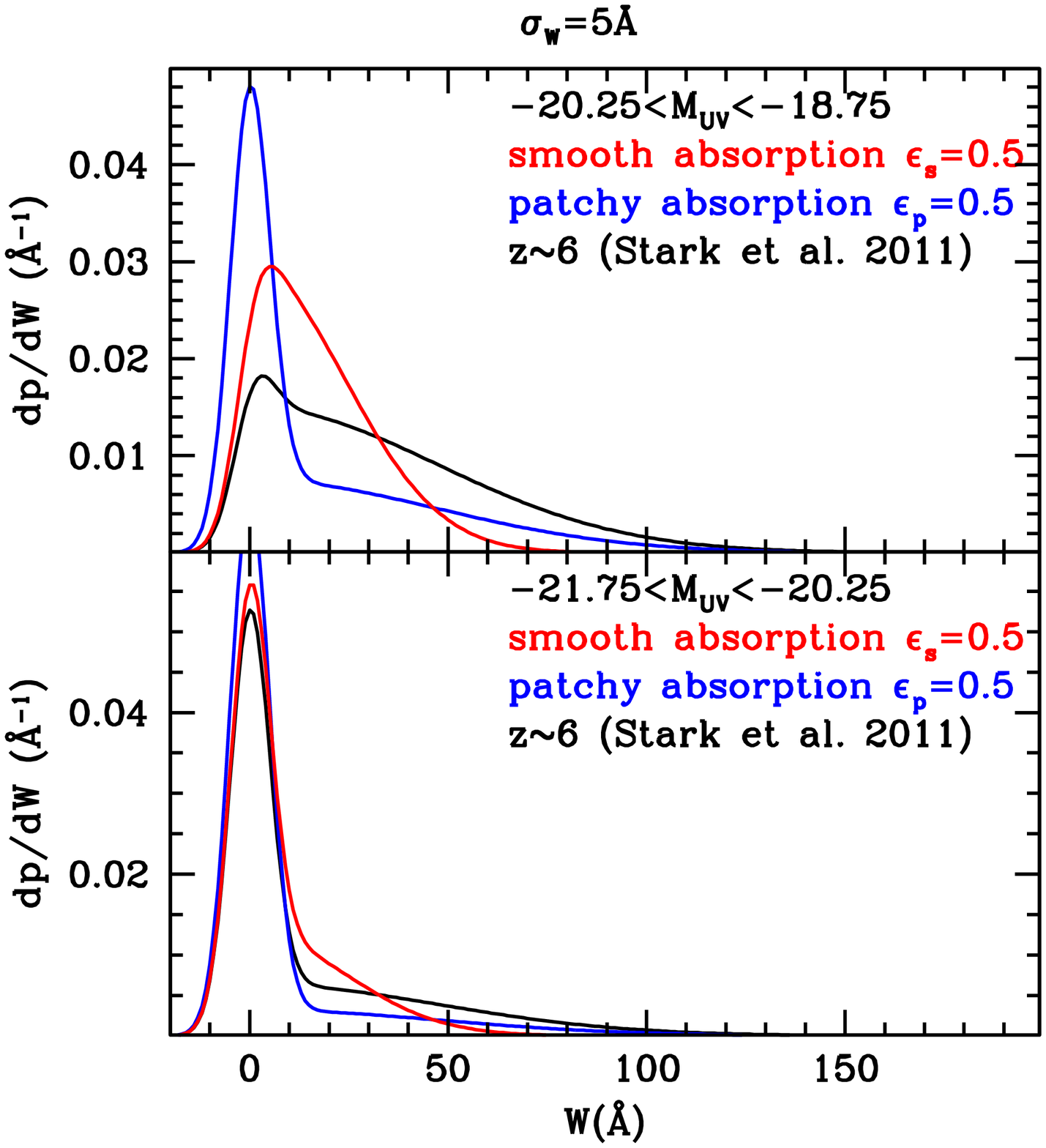}}
\end{center}
\figcaption{As in Figure~\ref{fig:models}, including a typical error of 5\AA\ on $W$. Non emitters now introduce a bump for small values of $W$, centered at zero.
\label{fig:modelse}}
\end{figure}

We consider two simple scenarios, illustrated
in~Figure~\ref{fig:models} and in Figure~\ref{fig:modelse} in the
presence of observational errors. The first, the patchy model, is
analogous to that considered by other authors
\citep{Fon++10,Sch++11,Pen++11,Ono++11} where a fraction of the
galaxies $\ep$ are completely absorbed (or do not emit at all, which
is equivalent in our model) while the remaining $1-\ep$ is
unabsorbed. In this case, the probability distribution at a higher
redshift than 6 is given by

\begin{equation}
p_{p}(W)=\ep p_6(W)+(1-\ep)\delta(W)= \frac{2A\ep}{\sqrt{2\pi}W_{c}} e^{-\frac{1}{2}\left(\frac{W}{W_{c}}\right)^2}H(W)+(1-A\ep)\delta(W).
\end{equation}
The second, the smooth model, assumes that all emission is
attenuated by a constant factor $\es$ so that
\begin{equation}
p_{s}(W)=p_6(W/\es)/\es= \frac{2A}{\sqrt{2\pi}\es W_{c}} e^{-\frac{1}{2}\left(\frac{W}{\es W_{c}}\right)^2}H(W)+(1-A)\delta(W).
\end{equation}

The two models describe in a very simple manner two interesting
physical scenarios, and illustrate the different strategies required
to investigate them. In the patchy model, there are overall fewer
emitters than in the smooth model, but they are found at higher
equivalent widths. For the examples shown in Figure~\ref{fig:models},
depending on the sensitivity one can find more sources in either
model: above $\sim$50\AA, one expects to find more sources in the
patchy case; below $\sim$50\AA\, the smooth model provides more sources.

\subsection{Application to spectroscopic data}

We now have to connect these distributions to the observables, a set
of fluxes measured at different wavelengths $\lambda_i$
\{$f_i=f(\lambda_i)$\}. For simplicity we consider 
an unresolved emission line, extracted without weighting from N$_{l}$
pixels, so that the effective noise is the noise measured within a
pixel multiplied by $\sqrt{N_l}$, while the effective flux is the flux
multiplied by $N_l$. Thus, the predicted flux is non zero only in the
pixel containing the redshifted Lyman $\alpha$ ($\lambda_0$) and it is
given by:

\be
f_p(\lambda=(1+z)\lambda_0)=W (1+z) f_o 10^{-0.4m}\frac{c}{\lambda_0^2(1+z)^2}\equiv Wf_m,
\label{eq:fp}
\ee
where $f_0=3.631\cdot 10^{-20}$ erg s$^{-1}$ Hz$^{-1}$ cm$^{-2}$, and
the first (1+z) transforms the rest frame equivalent width into
observer frame equivalent width. In order to take into account the
effects of resolution and line shape, and allow for optimal weighting,
it is sufficient to replace the above equation with an appropriate
function, e.g. a Gaussian of width equal to the resolution
$\sigma_\lambda$:

\be
f_p(\lambda)=\frac{Wf_m}{\sqrt{2\pi}\sigma_\lambda}e^{-\frac{1}{2}\left(\frac{\lambda-\lambda_0}{\sigma_\lambda}\right)^2}
\ee
Doing this correctly would require knowledge of the line profile, and
would add an additional convolution and unnecessary computational
burden at this stage. Therefore we will adopt the more conservative
approach outlined above and do not implement this refinement.

By combining Equations 1-4 with the appropriate Gaussian noise
\{$\sigma_i$\}, we can infer the posterior
probability of $\epsilon$ (which we use to indicate both $\ep$ and
$\es$) and $z$ given an observed spectrum and continuum magnitude
using Bayes' Theorem: 

\be
p(\epsilon,z_i|\{f\},m)=\frac{1}{Z}\left(\Pi_i \int dW p(f_i,m|W,z_i)p(W|\epsilon)\right) p(\epsilon)p(z_i),
\ee
i.e.:

\be
p(\epsilon,z_i|\{f\},m)=\frac{1}{Z}\int_0^{\infty} dW \left(\frac{1}{\sqrt{2
\pi}\sigma_i}e^{-\frac{1}{2}\left(\frac{f_i-Wf_m}{\sigma_i}\right)^2}\right) \Pi_{j\neq i} N(f_j,\sigma_j^2)  p(W|\epsilon) p(\epsilon)p(z_i)
\ee
where $z_i=\lambda_i/\lambda_0-1$, and $N(f_j,\sigma_j^2)$ is the
standard Gaussian (normal) distribution with mean $f_j$ and standard
deviation $\sigma_j$.  The likelihood is as usual the probability of
obtaining the data for any given value of the parameters
$p(\{f\},m|\epsilon,z_i)=\Pi_i p(f_i,m|\epsilon,z_i)$, and for
simplicity the error on $m$ has been considered negligible. For
simplicity we consider independent priors for $\epsilon$ and $z_i$,
even though one could easily implement a physically motivated prior,
where $\epsilon$ depends on $z_i$ (i.e. of the form
$p(\epsilon|z_i)p(z_i)$). The prior p($\epsilon$) is assumed to be
uniform between zero and unity, i.e. the intensity of Lyman$\alpha$
cannot increase beyond $z\sim6$; alternatively one could assume it to
be uniform between zero and 1/$A$, which is the maximum value
consistent with a probability density function positive
everywhere. The prior p($z_i$) is given by the photometric
redshift. Note that in the case of incomplete wavelength coverage
where p($z_i$) is non zero, our formalism will take this into account
correctly in deriving limits on $\epsilon$ and $z_i$.

The normalization constant $Z$ is known as the Bayesian Evidence and
quantifies how well the model matches the data. The evidence ratio is
a powerful way to perform model selection (e.g. comparing the patchy
and smooth models).  For a sample of galaxies, for multiple spectra of
the same galaxy, the likelihood is just the product of the individual
likelihoods, allowing for efficient combination of data of different
depths.

Considering the two specific models, the posterior distributions can
be derived analytically:

\be
p_p(\ep,z_i|\{f,\sigma\},m)=\frac{C}{Z}\left(\frac{A \ep \sigma_i \left(1+\rm{erf}(t_{m,p,i})\right)e^{-\frac{1}{2}\left[\left(\frac{f_i}{\sigma_{t,p,i}}\right)^2-\left(\frac{f_i}{\sigma_i}\right)^2\right]}}{\sigma_{t,p,i}}+(1-A\ep)\right)p(z_i),
\label{eq:ppz}
\ee
where
\be
C\equiv\Pi_j\frac{1}{\sqrt{2 \pi} \sigma_j}e^{-\frac{1}{2}\left(\frac{f_j}{\sigma_j}\right)^2},
\ee
is a constant depending only on the dataset, 
\be
\sigma_{t,p,i}\equiv\sqrt{\sigma_i^2+f_m^2(\lambda_i)W_c^2},
\ee
and
\be
t_{m,p,i}\equiv \frac{f_m(\lambda_i) W_c f_i}{\sqrt{2}\sigma_i \sigma_{t,p,i}}.
\ee
In the smooth case, the posterior probability distribution function is given by:
\be
p_s(\ep,z_i|\{f,\sigma\},m)=\frac{C}{Z}\left(\frac{\sigma_i A \left(1+\rm{erf}(t_{m,s,i})\right)e^{-\frac{1}{2}\left[\left(\frac{f_i}{\sigma_{t,s,i}}\right)^2-\left(\frac{f_i}{\sigma_i}\right)^2\right]}}{\sigma_{t,s,i}}+(1-A) \right)p(z_i),
\label{eq:psz}
\ee
where
\be
\sigma_{t,s,i}\equiv\sqrt{\sigma_i^2+f_m^2(\lambda_i)W_c^2\es^2},
\ee
and
\be
t_{m,s,i}\equiv \frac{f_m(\lambda_i) W_c f_i \es}{\sqrt{2}\sigma_i \sigma_{t,s,i}}.
\ee
In the patchy case the posterior pdf is separable and can be
integrated analytically to give the posterior pdf for the redshift
$z_i$

\be
p(z_i|\{f,\sigma\},m)=\int_0^1 d\ep p(\ep,z_i|\{f\},m)=
\ee

\be
=\frac{C}{Z}\left[\frac{A \sigma_i \left(1+\rm{erf}(t_{m,p,i})\right)e^{-\frac{1}{2}\left[\left(\frac{f_i}{\sigma_{t,p,i}}\right)^2-\left(\frac{f_i}{\sigma_i}\right)^2\right]}}{2\sigma_{t,p,i}}+\left(1-\frac{A}{2}\right)\right] p(z_i),
\ee

\begin{figure}
\begin{center}
\resizebox{0.49\columnwidth}{!}{\includegraphics{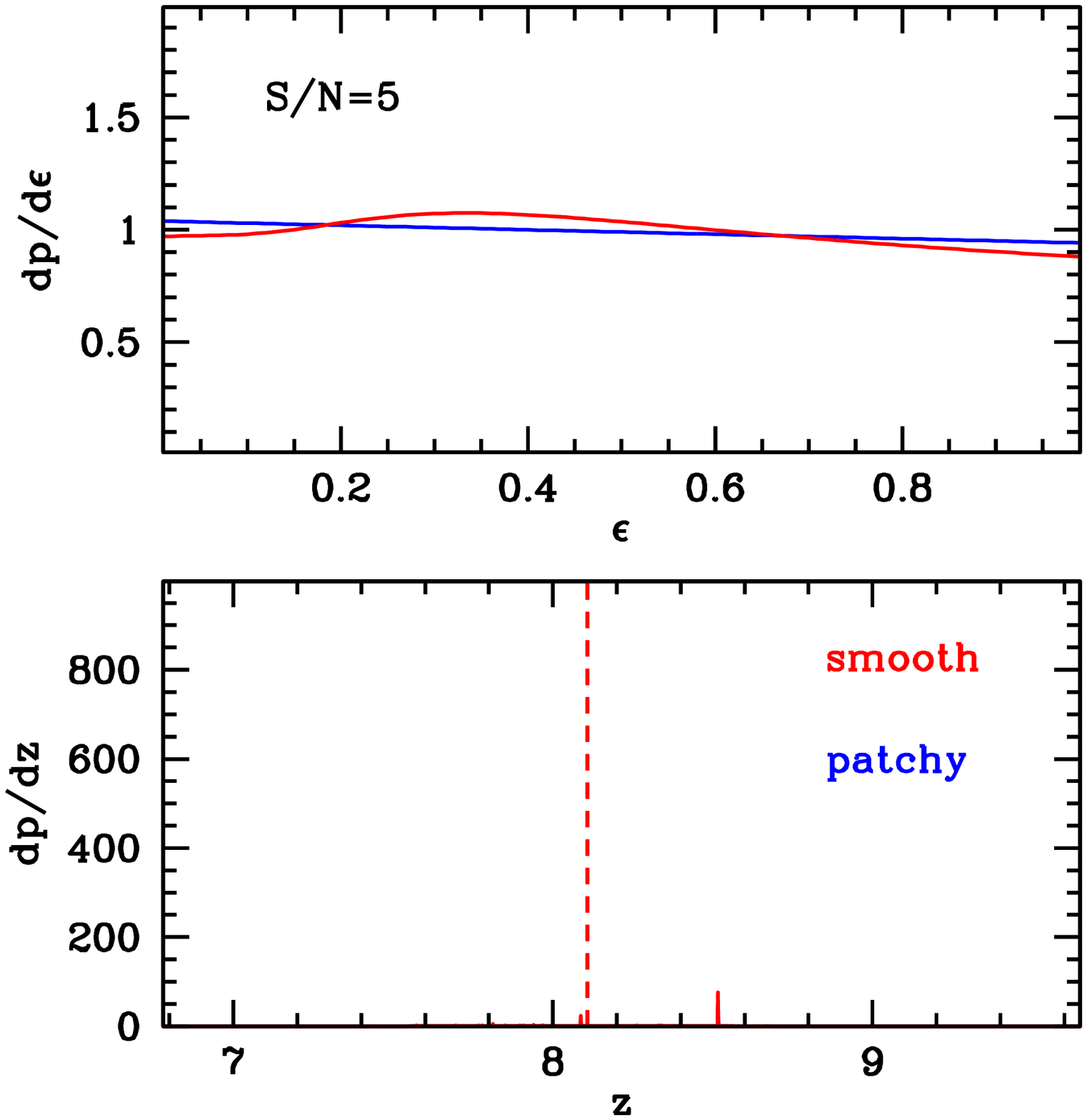}}
\resizebox{0.49\columnwidth}{!}{\includegraphics{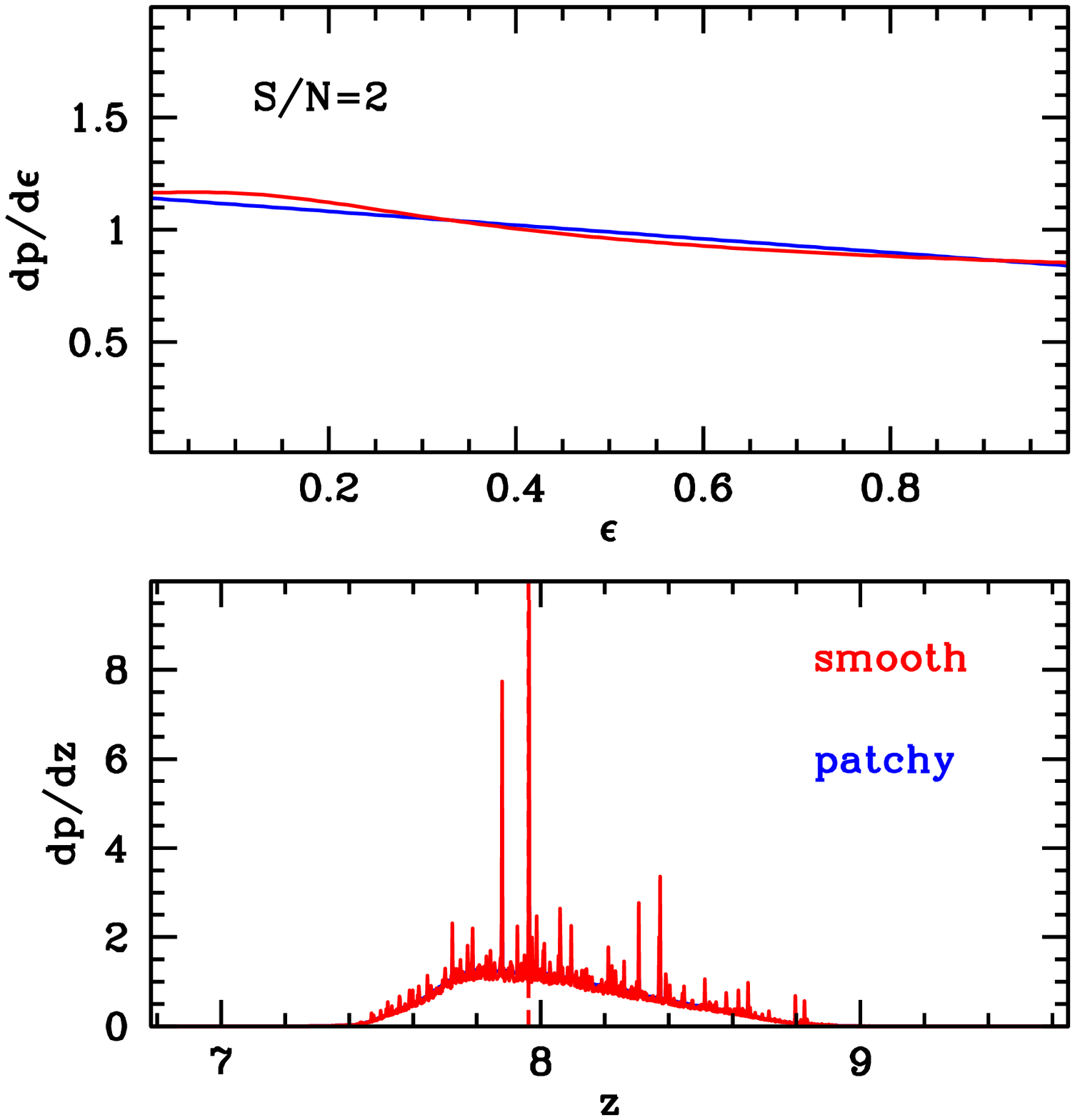}}
\end{center}
\figcaption{Illustration of inference on simulated data. For a 
simulated 5$-\sigma$ detection (left panels), the posterior pdf of $z$
(bottom panel) is sharply peaked at the redshift of the emission line
(vertical dashed line) independent of the adopted absorption model. As
expected, both models prefer large values of $\epsilon$ (top
panel). For a simulated weak (non) detection (2$-\sigma$; right
panels), there are insignificant noise peaks at several redshifts, on
top of the prior $p(z)$, consistent with the many $>$2-$\sigma$
fluctuations expected for a $\sim2000$ pixel spectrum. As expected,
both models prefer small values of $\epsilon$.
\label{fig:simul}}
\end{figure}

A simple illustration of this method applied to simulations is shown
in Figure~\ref{fig:simul}. Two emission lines with S/N=5 and S/N=2
have been added to noisy spectra covering the wavelength range
0.947-1.297 $\mu$m (equal to the range covered by our NIRSPEC
observations, described in Section~\ref{sec:obs}). We considered this
to be a bright galaxy and therefore used A=0.38 and
$W_c$=47\AA. Assuming a prior $p(z)$ appropriate for $Y$-band
dropouts, we computed the posterior pdf on $\epsilon$ and $z$. As can
be seen from the plots, the S/N=5 detection constrains the redshift
exquisitely well (vertical red dashed line), and tends to favor larger
values of $\epsilon$, i.e. emitters are common. The S/N=2 weak (non)
detection gives a posterior pdf with many spurious peaks in $z$, that
are not much higher than the prior distribution, consistent with the
fact that the likelihood of a false S/N$>$2 detection is large with
$\sim2000$ pixels. Conversely, since there are no strong lines, the
procedure correctly infers that $\epsilon$ should be small. Notice
that p($\epsilon$) is clearly non-Gaussian. With only one detection
not much can be learned about the distribution of $W$, and therefore
the posterior on $\epsilon$ is broad. In Sections~\ref{sec:results}
and~\ref{sec:predictions} we will consider more informative cases with
many sources.

In some cases, one might just wish to consider the inference on the
parameters if all it is known is that no line has been detected to a
certain level of significance, e.g. N$\sigma$. In this case, it is
sufficient to consider the integral of the likelihood, so that the
posterior pdf becomes in the patchy case:

\begin{multline}
p_p(\ep,z_i|\{\sigma\},N,m) \propto \frac{A \ep \{1+\rm{erf}[N\sigma_i/(\sqrt{2}\sigma_{t,p,i})]\}}{1+\rm{erf}(N/\sqrt{2})} \\  +\frac{2A\ep}{\sqrt{2\pi}[1+\rm{erf}(N/\sqrt{2})]}\int_{-\infty}^{N}dx_i \rm{erf}(t_{m,p,i})e^{-\frac{1}{2}\left(\frac{x_i \sigma_i}{\sigma_{t,p,i}}\right)^2}+(1-A\ep),
\end{multline}
where $t_{m,p,i}$ is only a function of the variable of integration
$x_i=f_i/\sigma_i$ and $\sigma_{t,p,i}$ does not depend on $f_i$. The
proportionality factor is $p(z_i)C_N/Z$, with
\be
C_N\equiv\Pi_j\int_{-\infty}^{N\sigma_j}\frac{1}{\sqrt{2 \pi} \sigma_j}e^{-\frac{1}{2}\left(\frac{f_j}{\sigma_j}\right)^2}=\left[\frac{\left(1+\rm{erf}(N/\sqrt{2})\right)}{2}\right]^{N_{\rm pix}},
\ee
for N$_{\rm pix}$ spectral pixels.

In the smooth case, the posterior pdf is given by:

\begin{multline}
p_s(\es,z_i|\{\sigma\},N,m) \propto \frac{A \{1+\rm{erf}[N\sigma_i/(\sqrt{2}\sigma_{t,s,i})]\}}{1+\rm{erf}(N/\sqrt{2})} \\  +\frac{2A}{\sqrt{2\pi}[1+\rm{erf}(N/\sqrt{2})]}\int_{-\infty}^{N}dx_i \rm{erf}(t_{m,s,i})e^{-\frac{1}{2}\left(\frac{x_i \sigma_i}{\sigma_{t,s,i}}\right)^2}+(1-A),
\end{multline}
where, again, $t_{m,p,i}$ is only a function of the variable of
integration $x_i=f_i/\sigma_i$ and $\sigma_{t,s,i}$ does not depend on
$f_i$. The constant of proportionality is, as in the patchy case,
$p(z_i)C_N/Z$,

\subsection{Application to flux catalogs}

Often one can only analyze flux catalogs, for example in narrow band
searches, or when only noise levels and non-detections are reported by
spectroscopic studies. In this case it is useful to consider a
simplified treatment, that allows one to combine heterogeneous data in
an efficient way. This is achieved by switching from flux to $W$ and
by integrating away (marginalize over) the dependency on redshift. In
this way, for any detection of an equivalent width $W_o$ with noise
level $\sigma_{W}$ the likelihoods for the two models are:

\be
p_p(W_o|\ep)=\int_0^\infty dW \frac{1}{\sqrt{2 \pi} \sigma_W}e^{
-\frac{1}{2}\left(\frac{W_o-W}{\sigma_W}\right)^2} \left(\frac{2A\ep}{\sqrt{2 \pi}W_c}e^{-\frac{1}{2}\left(\frac{W}{W_c}\right)^2}+(1-A\ep)\delta(W)\right)
\ee

\be
p_s(W_o|\es)=\int_0^\infty dW \frac{1}{\sqrt{2 \pi} \sigma_W}e^{
-\frac{1}{2}\left(\frac{W_o-W}{\sigma_W}\right)^2} \left(\frac{2A}{\sqrt{2 \pi}\es W_c}e^{-\frac{1}{2}\left(\frac{W}{\es W_c}\right)^2}+(1-A)\delta(W)\right).
\ee

As in the previous section, the integrals and posterior can be
computed analytically. In the patchy case the posterior is given by:

\be
p_p(\ep|W_o)=\frac{1}{Z}\left(\frac{A \ep \left(1+\rm{erf}(t_{m,p})\right)e^{-\frac{1}{2}\left(\frac{W_o}{\sigma_{W,t,p}}\right)^2}}{\sqrt{2\pi}\sigma_{W,t,p}}+
\frac{(1-A\ep)e^{-\frac{1}{2}\left(\frac{W_o}{\sigma_W}\right)^2}}{\sqrt{2\pi}\sigma_W}\right)p(\ep),
\label{eq:ppW}
\ee
where
\be
\sigma_{W,t,p}\equiv\sqrt{\sigma_W^2+W_c^2}
\ee
and
\be
t_{m,p}\equiv \frac{W_c W_o}{\sqrt{2}\sigma \sigma_{W,t,p}}.
\ee
In the smooth case, the posterior is given by:
\be
p_s(\es|W_o)=\frac{1}{Z}\left(\frac{A \left(1+\rm{erf}(t_{W,m,s})\right)e^{-\frac{1}{2}\left(\frac{W_o}{\sigma_{W,t,s}}\right)^2}}{\sqrt{2\pi}\sigma_{W,t,s}}+
\frac{(1-A)e^{-\frac{1}{2}\left(\frac{W_o}{\sigma_W}\right)^2}}{\sqrt{2\pi}\sigma_W} \right)p(\es),
\label{eq:psW}
\ee
where
\be
\sigma_{W,t,s}\equiv\sqrt{\sigma_W^2+W_c^2\es^2}
\ee
and
\be
t_{W,m,s}\equiv \frac{W_o W_c \es}{\sqrt{2}\sigma \sigma_{W,t,s}}.
\ee

If the only information available is about a subset of the wavelength
range where the line could possibly be found based on the photometric
redshift distribution, this can easily be implemented in this
formalism. Assuming for example that the line can be seen only in a
range between $z_{\rm min}$ and $z_{\rm max}$, in the patchy case the
expression is

\begin{multline}
p_p(\ep|W_o)\propto \left(\frac{A \ep \left(1+\rm{erf}(t_{m,p})\right)e^{-\frac{1}{2}\left(\frac{W_o}{\sigma_{W,t,p}}\right)^2}}{\sqrt{2\pi}\sigma_{W,t,p}}
\frac{(1-A\ep)e^{-\frac{1}{2}\left(\frac{W_o}{\sigma_W}\right)^2}}{\sqrt{2\pi}\sigma_W}\right)p(z\in [z_{\rm min},z_{\rm max}]) \\ +\frac{e^{-\frac{1}{2}\left(\frac{W_o}{\sigma_W}\right)^2}}{\sqrt{2\pi}\sigma_W} p(z\notin [z_{\rm min},z_{\rm max}]),
\label{eq:ppWz}
\end{multline}
where the constant of proportionality is $p(\ep)/Z$. A similar
expression applies for the smooth model.

As in the spectroscopic case, for non detections to a certain
noise level (e.g. $N \sigma_W$) the likelihood is just the integral of
the likelihood:

\be
p(W_o<N\sigma_W|\es)=\int_{-\infty}^{N\sigma_W} dW_o p(W_o|\es)
\ee

\begin{figure}
\begin{center}
\resizebox{0.97\columnwidth}{!}{\includegraphics{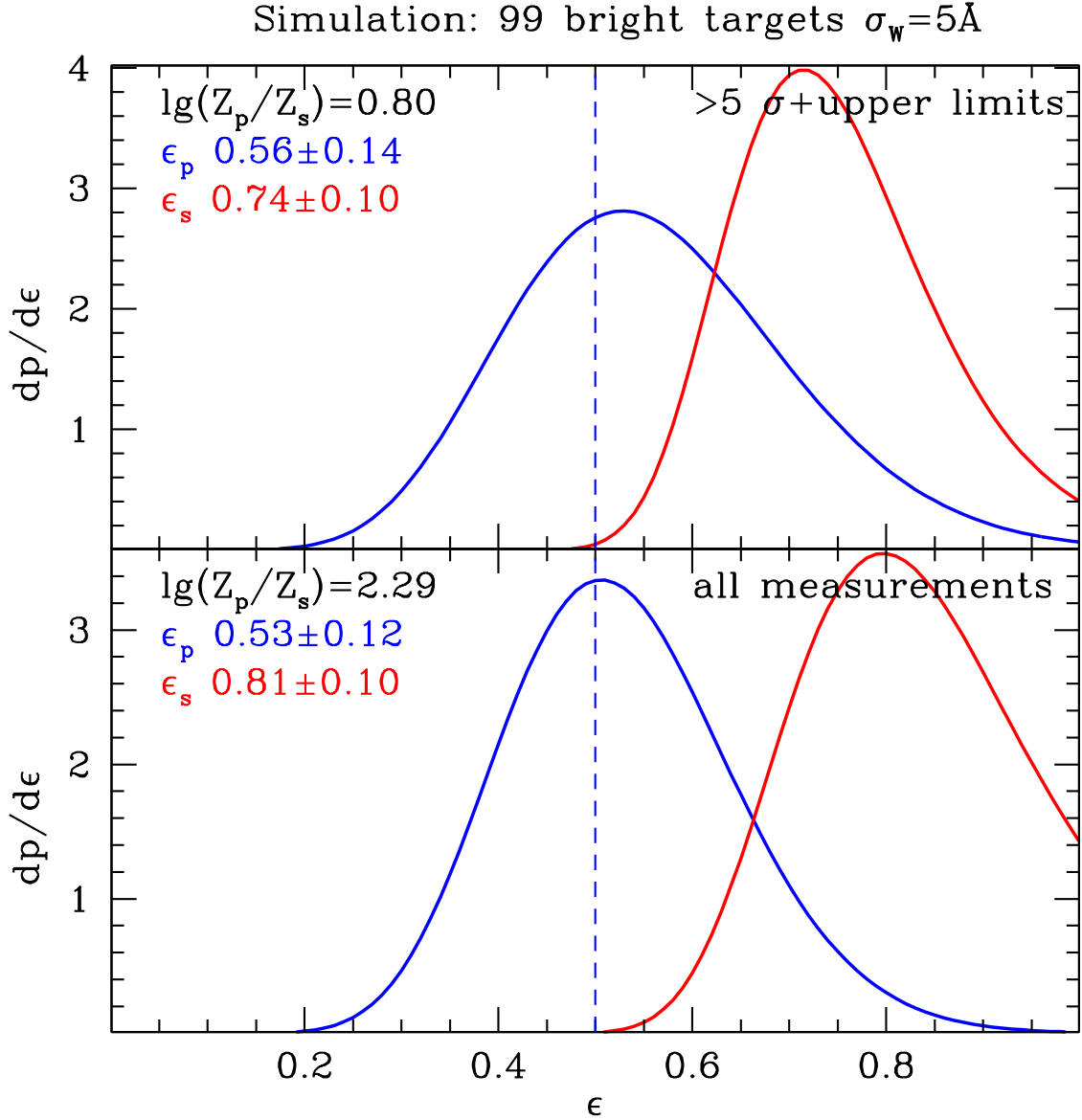}}
\end{center}
\figcaption{Inference on $\epsilon$ from a simulated sample of 99 objects at 
$z\sim 8$ with M$_{\rm UV}<-20.25$. The sample is generated from a
distribution of equivalent width with $\ep=0.5$, i.e. equal to that
shown in Figure~\ref{fig:models}, assuming a noise level equivalent to
5\AA equivalent width, i.e. 5-$\sigma$ detection limit of 25\AA.  The
top panel illustrates the inference based on counting non-detections
and measurements above 5-$\sigma$. The bottom panel utilizes all the
information, including non-detections. Both experiments recover the
correct value of $\ep$ and strongly prefer the patchy model (by a
factor of 6:1 in the upper panel and by 200:1 in the bottom
panel). Notice how the inferred value of $\es$, and the underlying
distribution, are dramatically different from the input, illustrating
the effects of using the wrong model to interpret the data, without
considering evidence for model selection.
\label{fig:simulWsp}}
\end{figure}

We illustrate the difference between using measurements and upper
limits only by means of simulations in Figure~\ref{fig:simulWsp}. We
construct a simulated dataset of 99 galaxies drawn from a distribution
with $\ep=0.5$, assuming noise $\sigma_W=5$\AA. In the top panel we
perform the inference based only on the detections with significance
5$-\sigma$ or more and counting the other objects as non-detections
(using the likelihood in Equation~29). In the bottom panel we used all
the available information from the full distribution of measured $W$
(using the likelihoods in Equation~20 and 21 even for fluxes below
5-$\sigma$). In both cases the inference accurately recovers the
correct value of $\ep$ and the evidence ratio selects patchy
absorption as the best model. However, uncertainties are marginally
smaller, and evidence ratio is much more conclusive when one utilizes
the full distribution. This underscores the importance of reporting
even marginal detections, if possible and if the errors are very well
known. If that is not possible, a careful treatment of upper limits is
still possible within this framework, and accurate.

\section{New observations}
\label{sec:obs}

\subsection{HST photometry and target selection}

The photometric data considered in this paper have been obtained as
part of the Hubble program GTO/COS 11534 (PI Green), retrieved from
the HST archive after the end of their proprietary period. Coordinated
parallels in six WFC3 filters where scheduled: three in the UVIS
channel (F300X: $4000$ s, F475X: $3200$ s, F600LP $3200$ s) and three
in near-IR (F098M: $17229$ s, F125W: $2006$ s, F160W: $2806$ s). The
program used the filter set of the HIPPIES survey \citep{Yan++11} with
the addition of F300X and F475X. Compared to the optimized exposure
times of our BoRG survey \citep{BORG1}, this set of observations has
an integration time in F098M that is about four times longer than
necessary to search for $z\sim 8$ galaxies given the depth of the $J$
and $H$ band exposures. Observations were not dithered.

We processed the data using our BoRG pipeline \citep{BORG1,BORG2}. We
calibrated individual exposures with calwfc3, then aligned and
registered them on a common $0\farcs08$/pixel scale using multidrizzle
\citep{Koe++02}. Sources were detected in the $J$-band image using 
SExtractor in dual image mode \citep{B+A96}, setting a threshold of at
least 9 contiguous pixels with $S/N>0.7$ after normalization of the
r.m.s. maps to take into account correlated noise \citep{BORG1}.

To select $z\sim 8$ candidates we require $S/N>8$ for ISOMAG flux in
the detection band ($J$) and $S/N>5$ in $H$ (ISOMAG). The standard
BoRG near-IR color-color selection has been applied:
\begin{equation}
Y-J \geq 1.75
\end{equation}
\begin{equation}\label{eq:J-H}
J-H<0.02+0.15\times(Y-H-1.75).
\end{equation}
Finally, we require a conservative non-detection in all three optical
bands ($S/N<1.5$) for ISOMAG fluxes. Flux measurements are corrected
for foreground Galactic extinction using the maps by \citet{SFD98},
which reports $A_B=0.29$ for the coordinates of the field. Colors are
measured using ISOMAG fluxes and have been PSF matched using the
latest WFC3 PSF
\citep[\url{http://www.stsci.edu/hst/wfc3/ir\_ee\_model\_smov.dat};
see also][]{BORG2}.

\begin{figure}
\begin{center}
\resizebox{0.75\columnwidth}{!}{\includegraphics{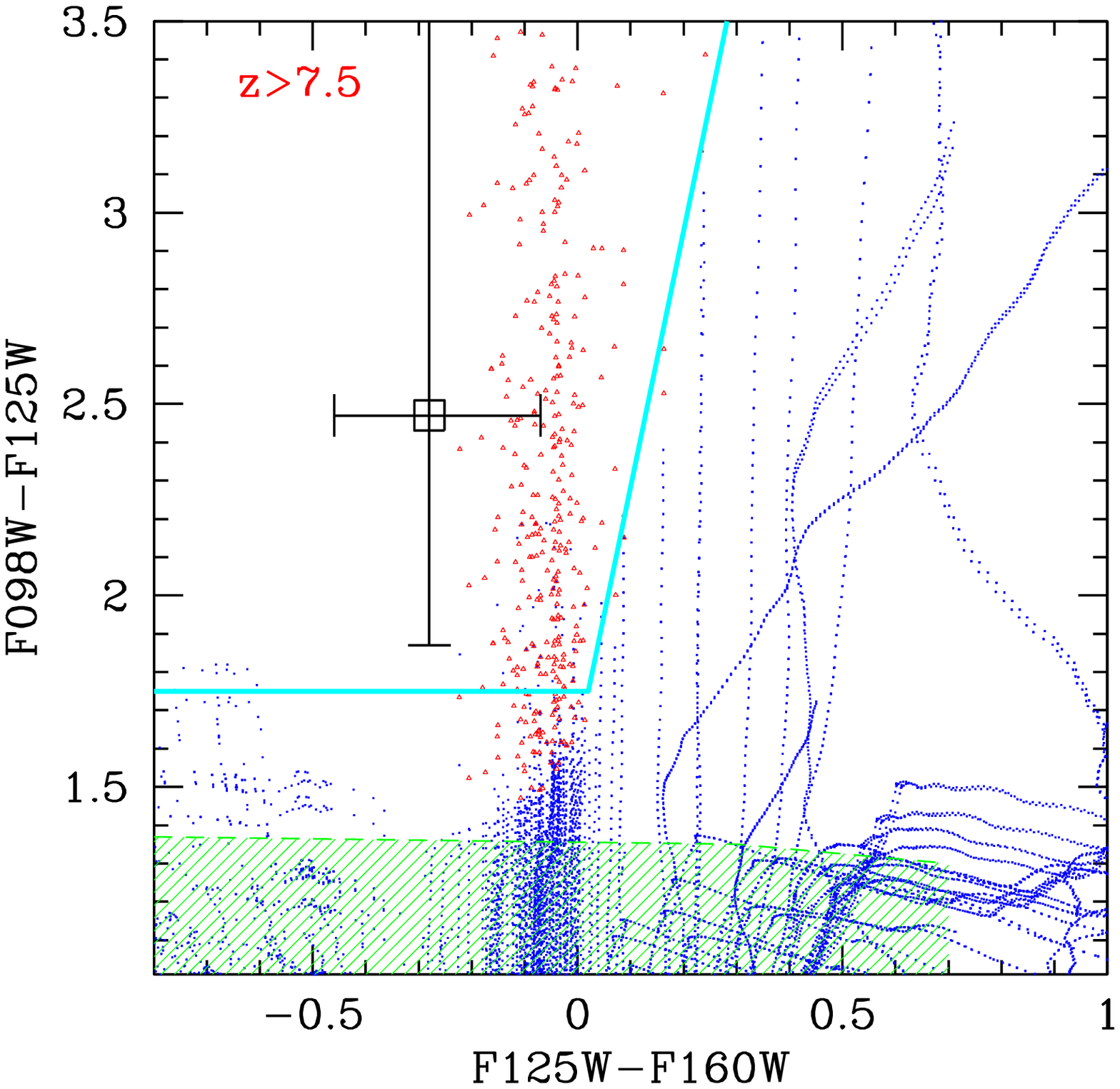}}
\resizebox{0.75\columnwidth}{!}{\includegraphics{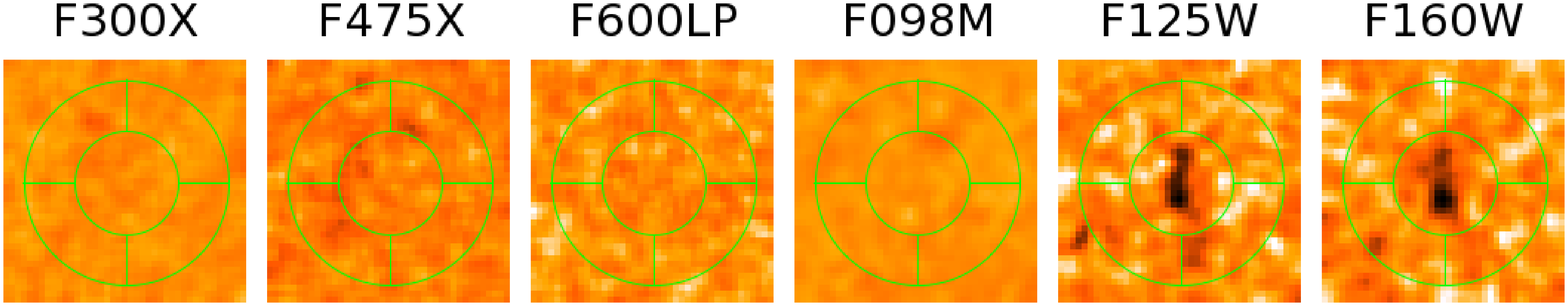}}  
\end{center}
\caption{Top panel: Color-color selection for $z\sim8$ candidates in
 the BoRG survey (from Trenti et al. 2011a). Red triangles are $z>7.5$
 simulated galaxies. Lower redshift contaminants are shown as blue
 dots (galaxies) and green region (L and T dwarf stars). BoRG11534
 (black square with errorbars; ISOMAG colors) is located on the track
 of $z>7.5$ sources and is well separated from possible contaminants.
 Lower panels: postage-stamps images ($3\farcs2\times 3\farcs2$) in the
 F300X, F475X, F600LP, F098M, F125W, and F160W bands of BoRG11534 from
 HST/WFC3 data.}
\label{fig:borg11534_wfc3}
\end{figure}

One source, located at coordinates 22:02:46.33 +18:51:29.5 (J2000)
satisfies our selection within the WFC3 field analyzed here. Its
photometry is summarized in Table~\ref{tab:photo}. The source is
detected with S/N=9.6 in $J$ and S/N=6.9 in $H$ (ISOMAG fluxes) and
has a marginal detection in the very deep $Y$ band data (S/N=1.5),
with a very red color in the Lyman break filters:
$Y-J=2.44^{+1.21}_{-0.60}$ (See Figure~\ref{fig:borg11534_wfc3}). The
source is clearly resolved in the F125W and F160W images
(Figure~\ref{fig:borg11534_wfc3}), ruling out contamination by a
foreground star.

\begin{deluxetable}{lrrr}
\tablecolumns{4} \tablewidth{0pt} \tablecaption{Photometry of $z\sim
8$ candidate.}\label{tab:photometry}

\tablehead{\colhead{Filter} & \colhead{mag$_{ISO}$} &
\colhead{mag$_{FIXED}$} & \colhead{mag$_{AUTO}$} }
\startdata 
F160W &       $26.29 \pm 0.15$ &  $26.51 \pm 0.16$ &  $25.86  \pm 0.16 $ \\
F125W &       $26.01 \pm 0.11$ &  $26.54 \pm 0.15$ &  $25.98 \pm 0.16$ \\   
F098M &       $28.45^{+1.15}_{-0.55}$ &  $>29.04$ & $27.97^{+1.17}_{-0.56}$ \\
F600LP &      $>28.46$   &        $>28.42$ & $>27.85$ \\
F475X  &      $>28.42$   &        $>28.56$ & $>27.82$ \\
F300X  &      $>27.07$   &        $>27.21$ & $>26.51$ \\
\enddata
\tablecomments{Photometry for the $z\sim 8$ galaxy candidate discussed
in the paper. First column: filter. Second column ISOMAG magnitude,
with error. Third column: magnitude within a fixed aperture of radius
$r=0\farcs32$. Fourth column: total magnitude (AUTOMAG). ISOMAG and
fixed aperture measurements have been PSF matched to the $J$ band. All
measurements have been corrected for galactic reddening using
extinction as measured by \citet{SFD98}.\label{tab:photo}}
\end{deluxetable}

\subsection{Keck Spectroscopy}

\begin{figure}
\begin{center}
\resizebox{\columnwidth}{!}{\includegraphics{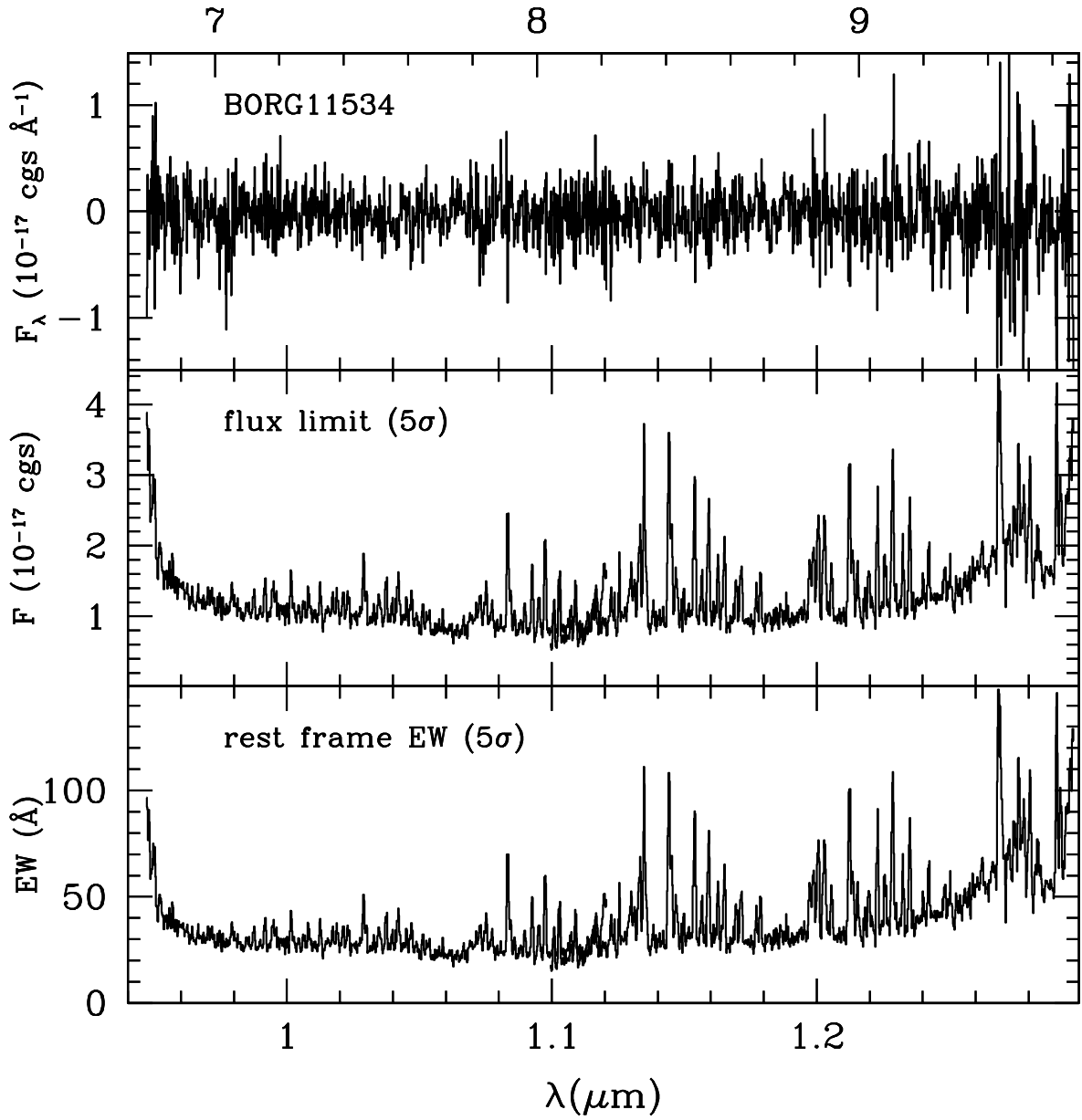}}
\end{center}
\figcaption{Keck spectroscopy of $Y$-band dropout BoRG11534. The top panel shows the measured spectrum; the middle panel shows the equivalent 5-$\sigma$ flux limit for an unresolved line; the bottom panel shows the corresponding 5-$\sigma$ equivalent width limit.
\label{fig:spec}}
\end{figure}

The $Y$-band dropout BoRG11534 (Fig~\ref{fig:spec}) was observed using
the NIRSPEC spectrograph \citep{McL++98} on the night of August 13
2011.  The seeing was excellent ($0\farcs4$-$0\farcs5$), and even
though part of the night was lost to fog and clouds, we were able to
observe the target for 2.5 hrs each in the N1 and N2 setup, covering
the wavelength interval 0.9470-1.2969 $\mu$m, corresponding to Lyman
$\alpha$ redshifted to z=6.78-9.66, i.e. the range expected for
$Y$-band dropouts.

A bright star ($J_{125}=17.78$) was observed in the slit together with
the dropout in order to ensure optimal extraction and thus maximize
sensitivity, and to provide a secondary spectrophotometric standard,
identified as a rK4III star by comparing its colors to those predicted
by the \citet{Pic98} spectral library (Java applet available at \url{
http://lcogt.net/ajp/SpecMatch/hst}).

The data were reduced in a standard manner using a set of python
scripts. The extracted 1-d spectrum and the noise spectrum are shown
in Figure~\ref{fig:spec}. No significant emission is detected. For
comparison with other work, we also derive the corresponding flux and
equivalent width 5-$\sigma$ limit for an unresolved emission line. The
median 5-$\sigma$ limits are $(0.98\pm0.17)\cdot 10^{-17}$ erg s$^{-1}$
cm$^{-2}$ and (26$\pm$4) \AA\ in the N1 filter, and $(1.10\pm0.3)\cdot
10^{-17}$ erg s$^{-1}$ cm$^{-2}$ and (35$\pm$11) \AA\ in the N2
filter. The error bars represent the 25 and 75 percentile intervals.

The non-detection sets one of the most stringent upper limits to the
equivalent width of emission lines for a $Y$-band dropout, and all but
rules out faint emission line objects at lower redshifts as a
potential contaminant, as in the case of the observations of BoRG58 by
\citet{Sch++11} discussed by
\citet{BORG2}. In fact, as suggested by \citet{Ate++11}, faint
emission line objects at appropriate redshifts ([\ion{O}{2}] and
[\ion{O}{3}] or [\ion{O}{3}]/H$\beta$ and H$\alpha$) could be mistaken
for $z\sim8$ galaxies when only two detection bands are
available. However, if the continuum magnitude in F125W were due to an
emission line, it would correspond to $\sim 8 \cdot 10^{-17}$ erg
s$^{-1}$cm$^{-2}$, easily detectable with our sensitivity and
wavelength coverage. The only exception would be if weak [\ion{O}{3}]
fell beyond 1.2969$\mu$m ($z>1.590$) but within the F125W filter. In
that case however, H$\alpha$ would fall at 1.6999$\mu$m, just outside
the F160W filter, and thus would be inconsistent with the detection in
$H$.

\section{Current limits}
\label{sec:results}

In \S~\ref{ssec:7} we apply our methodology to a compilation of
published systematic spectroscopic studies of $z'$-band dropouts
deriving robust constraints on the distribution of Lyman-$\alpha$
emission at $z\sim7$. Then, in~\ref{ssec:8} we show that existing
spectroscopic samples of $Y$-dropouts, including our new upper limit,
are not sufficient to constrain the distribution of Lyman-$\alpha$
emission at $z\sim8$.

\subsection{Inference from $z'$-band dropouts at $z\sim7$}
\label{ssec:7}

In order to obtain an unbiased estimate it is essential to analyze
datasets for which detections and non-detections have both been
reported. The depth and observational configuration need not be the
same, but serendipitous discoveries are difficult to incorporate.  For
this reason we limit our analysis to three recently published samples
of $z'$-band dropouts, for which complete information is available
\citep{Pen++11,Sch++11,Ono++11}. The total sample consists of 39 
$z'$-band dropouts: 20 objects studied by \citet{Pen++11}, 11 objects
studied by \citet{Ono++11}, and the eight objects in the top part of
Table~1 of the paper by \citet{Sch++11} for which deep LRIS
spectroscopy is available. Aiming to ensure homogeneity in our
constraints at $z\sim7$ we do not include objects with estimated
photo-z above $z=7.5$, or objects for which only NIRSPEC coverage is
available. For each object we consider the appropriate measurements or
upper limits on line equivalent width as quoted by the authors, and we
use the parameters $A$ and $W_c$ appropriate for its absolute UV
magnitude.  

\begin{figure}
\begin{center}
\resizebox{0.97\columnwidth}{!}{\includegraphics{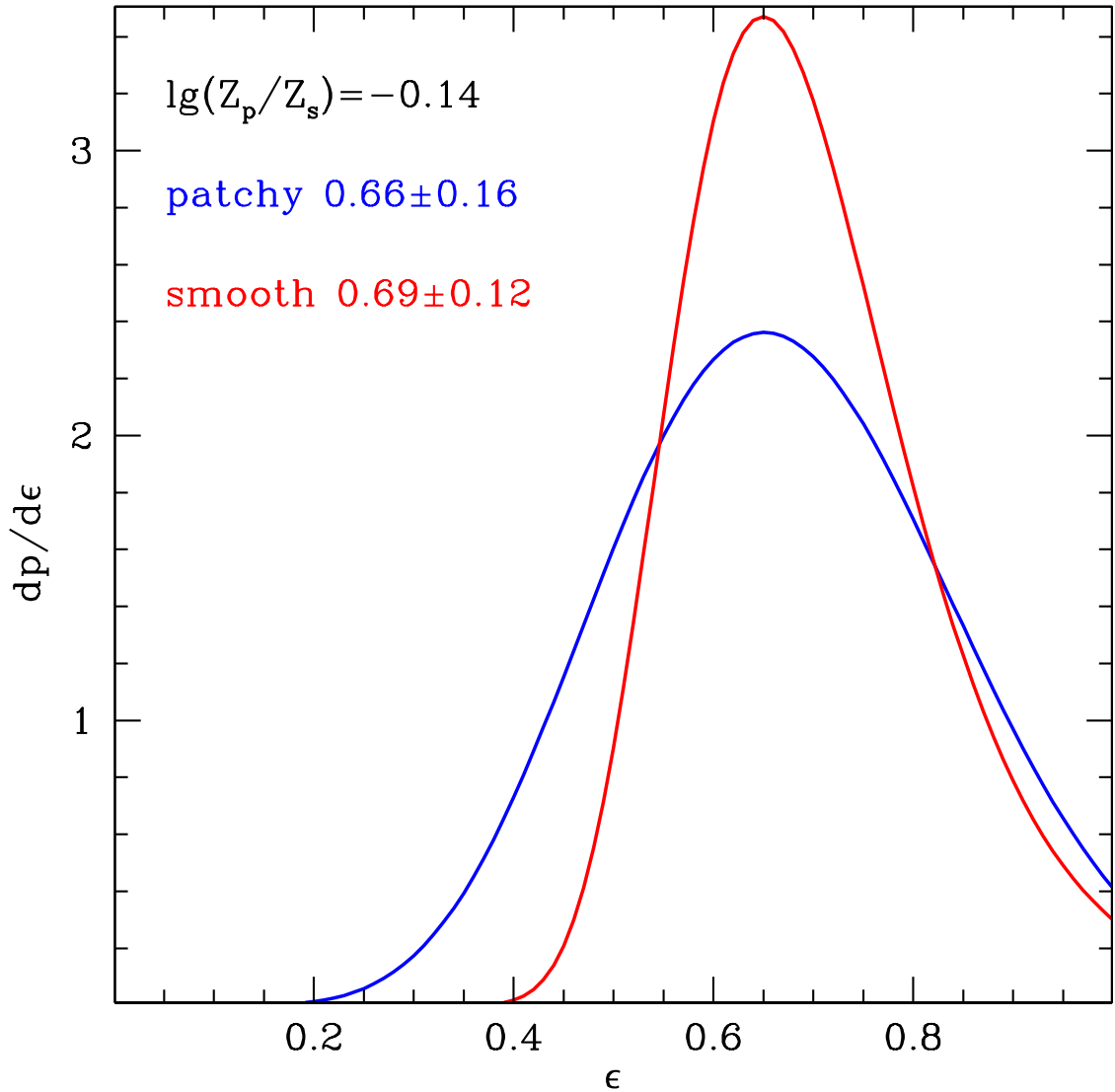}}
\end{center}
\figcaption{Marginalized posterior distribution function of $\epsilon$ at $z\sim7$ based on a compilation of 39 $z'$-dropouts with deep spectroscopic follow-up taken from the literature \citep{Pen++11,Ono++11,Sch++11}. Both the patchy and smooth model indicate clearly that the Lyman $\alpha$ emission is significantly quenched at $z\sim7$ with respect to $z\sim6$.
\label{fig:zdrops}}
\end{figure}

As shown in Figure~\ref{fig:zdrops} the data clearly prefer
$\epsilon<1$, independent of the model considered. The evidence ratio
indicates no significant preference for either model. For the patchy
model, we find $\ep=0.66\pm0.16$. For the smooth model we find
$\es=0.69\pm0.12$.  Our analysis gives consistent results, albeit with
larger errors and marginal differences, for each of the subsamples
(Ono $\ep=0.75\pm0.19$, $\es=0.74\pm0.15$; Pentericci
$\ep=0.59\pm0.18$, $\es=0.66\pm0.14$; Schenker $\ep=0.51\pm0.25$,
$\es=0.69\pm0.16$).

In terms of Gaussian approximation of the posterior, $\epsilon=1$ is
rejected at more than two standard deviations.  An increased fraction
of interlopers with respect to analogous samples at $z\sim6$, could
potentially explain this finding. However, assuming a typical fraction
of $\sim 25\%$ at $z\sim6$~\citep{Fon++10}, would require the fraction
of interlopers to be $\sim 50$\% at $z\sim7$, i.e. double. This seems
highly unlikely considering that the technique is the same and the
quality of the photometry is the same. We thus confirm the finding
that the distribution of Lyman $\alpha$ equivalent widths is
significantly weaker at $z\sim7$ with respect to $z\sim6$ possibly
signaling the onset of cosmic reionization \citep{Fon++10}.

We can give a simple interpretation of our results noticing that
$\epsilon$ corresponds to the average excess optical depth of
Lyman$\alpha$ with respect to $z\sim6$, i.e. $\langle e^{-\tau_{\rm
Ly\alpha}}\rangle$. Therefore our measurement implies $\langle
\tau_{\rm Ly\alpha} \rangle =0.4\pm0.2$. In order to interpret this number
correctly one cannot assume a uniform ionized medium \citep{Mir98},
but it is essential to take into account local HII regions, whose size
depends on the efficiency of galaxies in producing ionizing photons.
Furthermore, it is also essential to take into account clustering,
since nearby sources also contribute to the size of the ionized
region. In this scenario, we can use the models by
\citet{W+L05} to connect our observed optical depth to the fraction of
neutral hydrogen. The typical luminosity of M$_{\rm UV}\sim-20$ of the
$z\sim7$ sample, corresponds to a halo mass of $\sim 1.5\cdot 10^{11}$
M$_\odot$ \citep{Trenti++10}, and therefore a circular velocity of
$\sim170$~kms$^{-1}$, and velocity dispersion $\sim120$
kms$^{-1}$. Thus, our measured optical depth falls at the low end of
the range predicted by their models at $z\sim7$, i.e. consistent with
a ionized fraction of hydrogen of 0.4-0.7.  We note that this result
depends critically on the local environment of the galaxies rather
than on the average properties of the intergalactic medium, and
therefore our conclusions on the fraction of ionized gas should be
taken with a grain of salt.

\subsection{Inference from $Y$-band dropouts at $z\sim8$}
\label{ssec:8}

The situation is much less well-defined at $z\sim8$ and above. Few
reports of deep spectroscopic follow-up of WFC3-selected $Y$-band
dropouts are reported in the literature \citep{Sch++11}, owing to the
challenges of near infrared spectroscopy from the ground. Only one
detection has been reported to our knowledge by \citet{Leh++10}, and
with unusually high equivalent width, and significance just above the
conventional threshold ($S/N\sim6$). Therefore we do not expect our
inference to be conclusive, but nevertheless it is useful to
illustrate our current limits, in view of the future studies that we
will discuss in the next section.

\begin{figure}
\begin{center}
\resizebox{0.99\columnwidth}{!}{\includegraphics{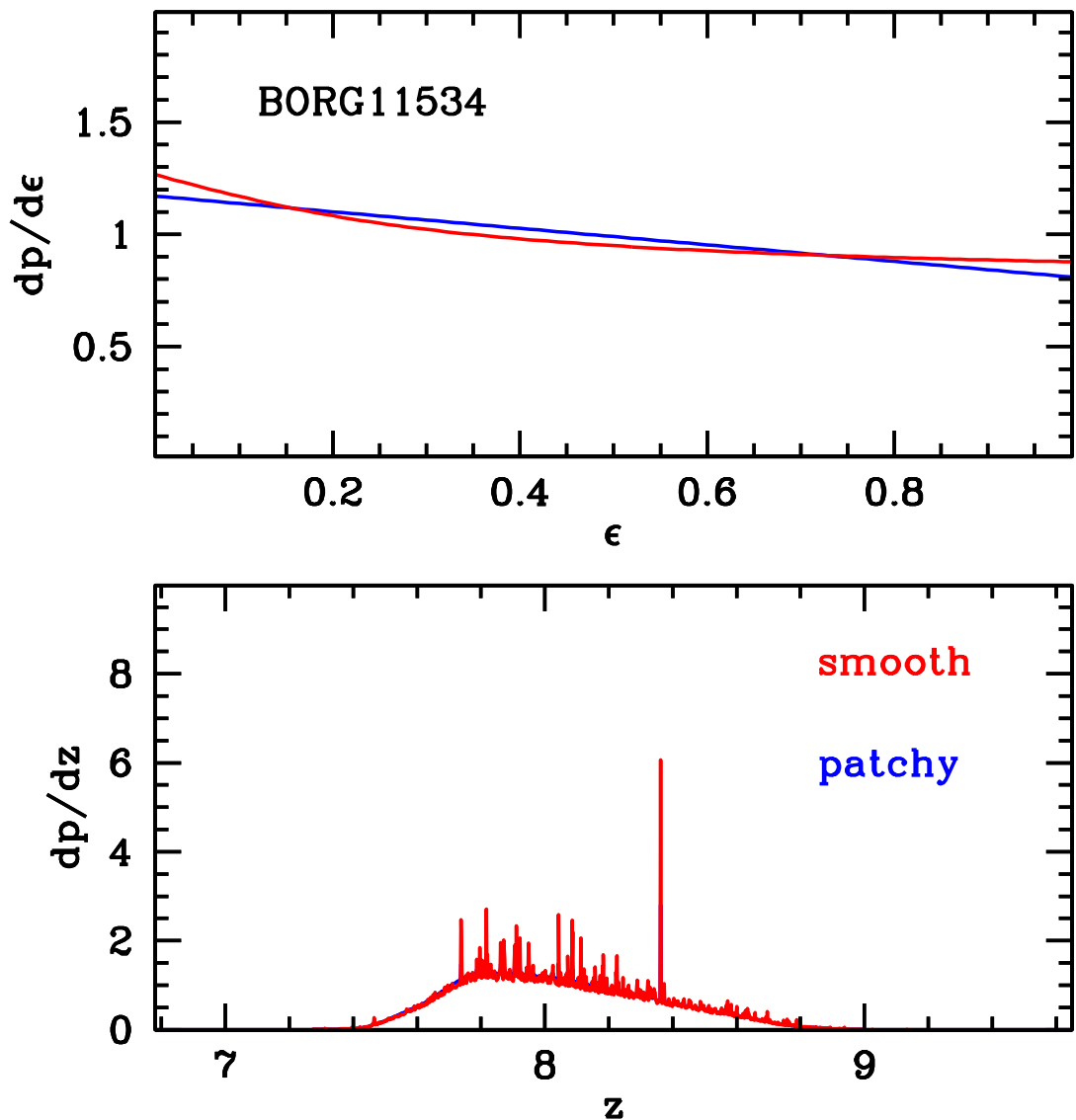}}
\end{center}
\figcaption{Marginalized posterior distribution functions based on the new observations of BoRG11534, presented in this paper. As expected, the non-detection implies no-constraints on the redshift (the small spike at $z\sim8.4$ is insignificant and fully expected given the number of pixels; see the right panel of Figure 3), while it implies a very weak preference for small values of $\epsilon$.
\label{fig:BORG}}
\end{figure}

We begin by analyzing our own non-detection of BoRG11534. For this
dataset we can exploit the full spectrum and take advantage of all the
available information. The marginalized posterior pdfs are shown in
Figure~\ref{fig:BORG}. As expected both the redshift and $\epsilon$
parameters are unconstrained by the data.

We therefore add the two $z\sim8$ objects from the sample of
\citet{Sch++11} that have complete wavelength coverage 
from NIRSPEC: A1703\_zD7 \citep{Bra++11} BoRG58 \citep[which is
selected from our BoRG survey;][]{BORG1}. Given the extreme faintness
of A1703\_zD7, even accounting for lensing magnification, most of the
constraints come from the two BoRG targets. To analyze these two
objects we use the version of the formalism developed for flux upper
limits, adopting the median equivalent width limit over the spectral
range, as estimated by scaling our observed noise to the actual
exposure time (the instrumental configuration is the same). As
expected, the data show a weak preference for small values of
$\epsilon$, although clearly the evidence is inconclusive. Note if the
fraction of bright lyman-$\alpha$ emitters at $z\sim6$ were higher
than in the sample published by~\citet{SEO11} as recently suggested by
\citet{Cur++11}, the preference for small values of $\epsilon$ would
increase, although not significantly given the small sample size at
$z\sim8$.

As a test we also add the detection of the object from
\citet{Leh++10}. Interestingly, consistent with the 
high equivalent width of the detection, the results are significantly
different for the two models. The smooth model is only capable of
producing the event for large values of $\epsilon$, while the patchy
model can explain the observations with a broader range of parameter
values. This is reflected in the evidence, which marginally prefers
the patchy model by a factor of 3:1 if the detection is included,
while the two models are indistinguishable if the detection is
excluded.

\begin{figure}
\begin{center}
\resizebox{0.97\columnwidth}{!}{\includegraphics{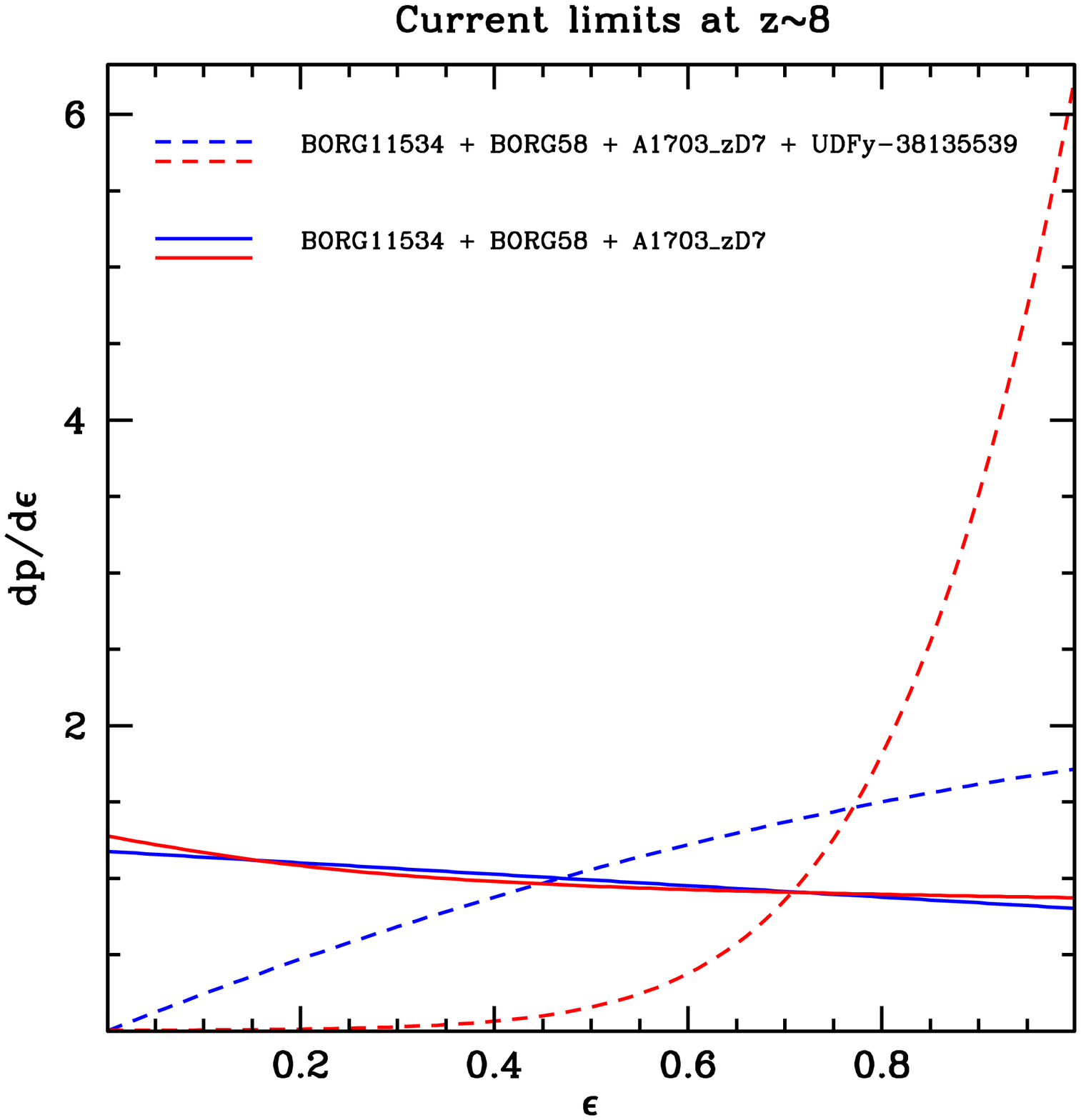}}
\end{center}
\figcaption{Marginalized posterior distribution function of $\epsilon$ based on the new observations of BoRG11534, as well as the study of three other objects from the literature. As expected, the non-detections imply a very weak preference for small values of $\epsilon$. The addition of the detection of UDFy-38135539 \citep{Leh++10} pushes the inference to larger values of $\epsilon$ the detection is more unlikely under the smooth model which therefore expresses a stronger preference for $\es\approx=1$, although overall the smooth model is marginally disfavored by the evidence.
\label{fig:ydrops}}
\end{figure}

\subsection{Comparison with previous work}

Our results at $z\sim7$ are based on the deep and comprehensive
optical follow-up of $z'$-dropouts performed by three groups
\citep{Sch++11,Ono++11,Van++11,Fon++10,Pen++11}. Although our 
methodology allows us to determine more than just the fraction of
emitters above a certain threshold it is straightforward to compare
with the commonly reported fraction of Lyman$\alpha$ emitters above a
certain threshold, typically $X^{55}$ and $X^{25}$.

In the patchy model the fraction of Lyman$\alpha$ emitters is simply
$X^W_{z=7}=\ep X^W_{z=6}$, where $X^W_{z=6}$ is the reference
measurement at $z=6$ (in this case taken from Stark et al.\ 2011). For
the bright subsample we find $X^{25}_{z=7}=\ep
(0.20\pm0.08)=0.14\pm0.06$ and $X^{55}_{z=7}=\ep
(0.074\pm0.050)=0.05\pm0.04$.  For the faint subsample we find
$X^{25}_{z=7}=\ep(0.54\pm0.11)=0.37\pm0.11$ and $X^{55}_{z=7}=\ep
(0.27\pm0.08)=0.19\pm0.07$.

In the smooth model the fraction of Lyman$\alpha$ emitters is
$X^W_{z=7}=\frac{\rm{erfc}(W/\sqrt{2} \es
W_c)}{\rm{erfc}(W/\sqrt{2}W_c)} X^W_{z=6}$. Thus, for the bright
subsample the smooth model implies $X^{25}_{z=7}=0.14\pm0.06$ and
$X^{55}_{z=7}=0.02\pm0.02$. For the faint subsample, the smooth model
implies $X^{25}_{z=7}=0.38\pm0.11$ and $X^{55}_{z=7}=0.08\pm0.03$.

We conclude that the models give mutually consistent emitter fractions
within the errors, except for $W>55$\AA, where by construction the
patchy model has significantly more probability. Below 25\AA\ the
converse would be true. More data are needed to distinguish the two
models as discussed in the previous section.

The agreement with published data is excellent for the patchy model,
which is equivalent to that implicitly assumed by previous
authors. However, the slightly different results for the smooth model
emphasize that it is important to recognize the inevitable underlying
model when analyzing data. Furthermore, our uncertainties are
significantly smaller than those quoted by \citet{Ono++11} using
virtually the same data, by virtue of our ability to take into account
strength and significance of non-detections, rather than just counting
detections above a certain threshold. More data are necessary to
determine which model is a better description of the data, including
of course more general models than the one discussed here.

Finally, we note that our interpretation of the findings in terms of
ionized fraction of neutral hydrogen is consistent with that of
\citet{Pen++11} based on the models by \citet{DMW11}.

\section{Forecasts}
\label{sec:predictions}

We conclude by presenting forecasts for observing campaigns of
$z\sim8$ galaxies. Given the paucity of strong emitters among the
dropout population it is clear that multiplexing capabilities, such as
those afforded by grism spectroscopy using the WFC3-IR channel, or
those available or soon to be available from the ground, will be key
to make progress. The question is what is the optimal strategy (how
deep and how many objects one has to observe) in order to make
progress, for example distinguishing the two empirical models
introduced in this paper. The simulations shown in
Figure~\ref{fig:simulWsp} give a first answer: by observing 99 objects
to the current best limiting depth it should be possible to answer the
question definitively. However, finding 99 bright $Y$-band dropouts
will require an order of magnitude more survey area than what is
planned to be observed so far with WFC3 and considerable effort for
follow-up, given their low density on the sky even with
clustering~\citep{BORG1,BORG2}.
  
\begin{figure}
\begin{center}
\resizebox{0.97\columnwidth}{!}{\includegraphics{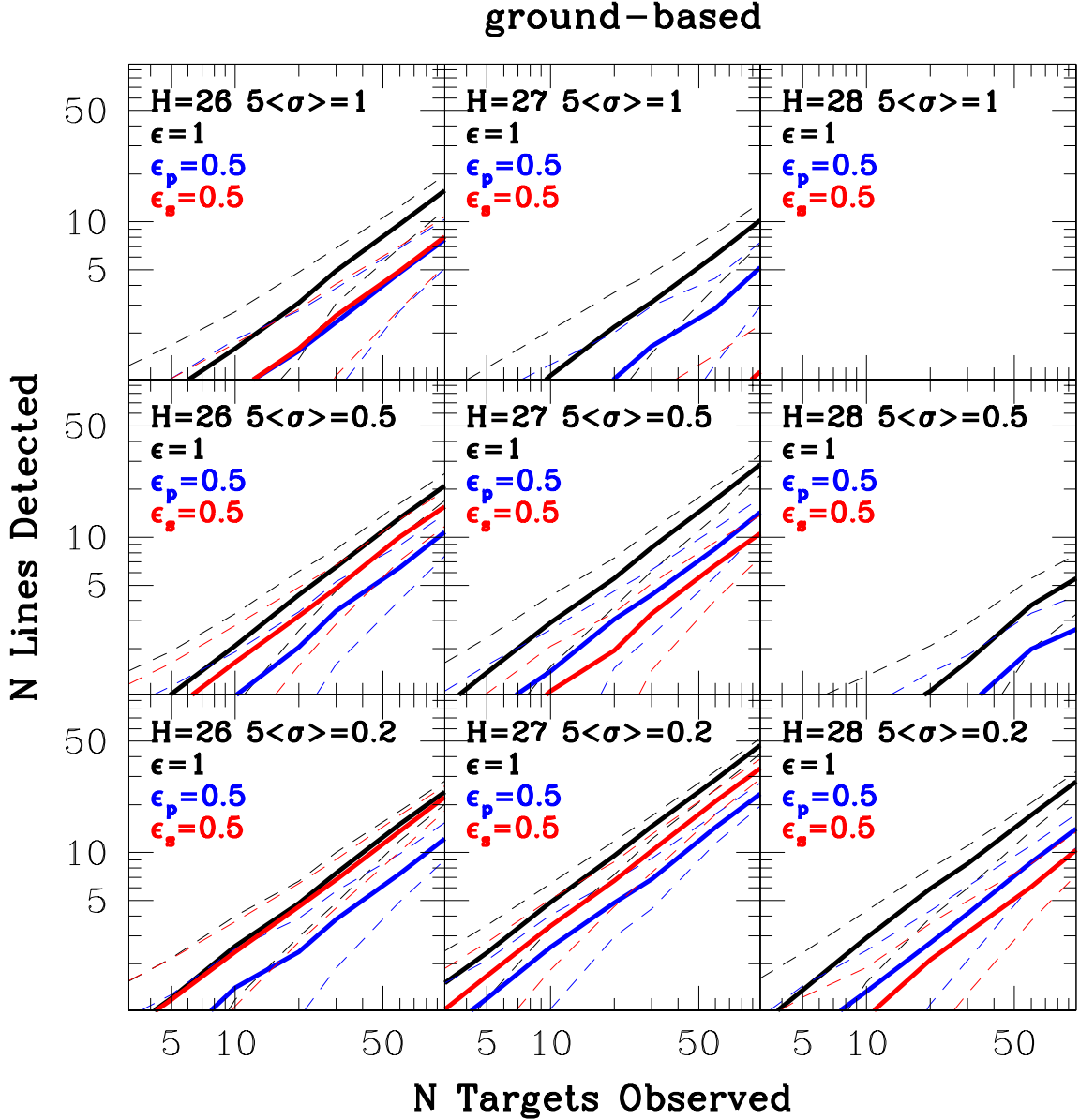}}
\end{center}
\figcaption{Predicted detection rates for ground based observations of Y-band dropouts as a function of continuum depth and spectroscopic sensitivity. Spectroscopic sensitivity is given in units of 10$^{-17}$ erg s$^{-1}$ cm$^{-2}$. Mean number of detections (solid line) and 1-$\sigma$ confidence contours (dashed lines) are shown for three reference models: 1) $\epsilon=1$ i.e. Lyman$\alpha$ distribution as at $z\sim6$ (black lines); 2) $\ep=0.5$ (blue lines), i.e. half the emitters as at $z\sim6$; 3) $\es=0.5$ (red lines), i.e. half the intensity of emission as at $z\sim6$.  
\label{fig:ground}}
\end{figure}

\begin{figure}
\begin{center}
\resizebox{0.97\columnwidth}{!}{\includegraphics{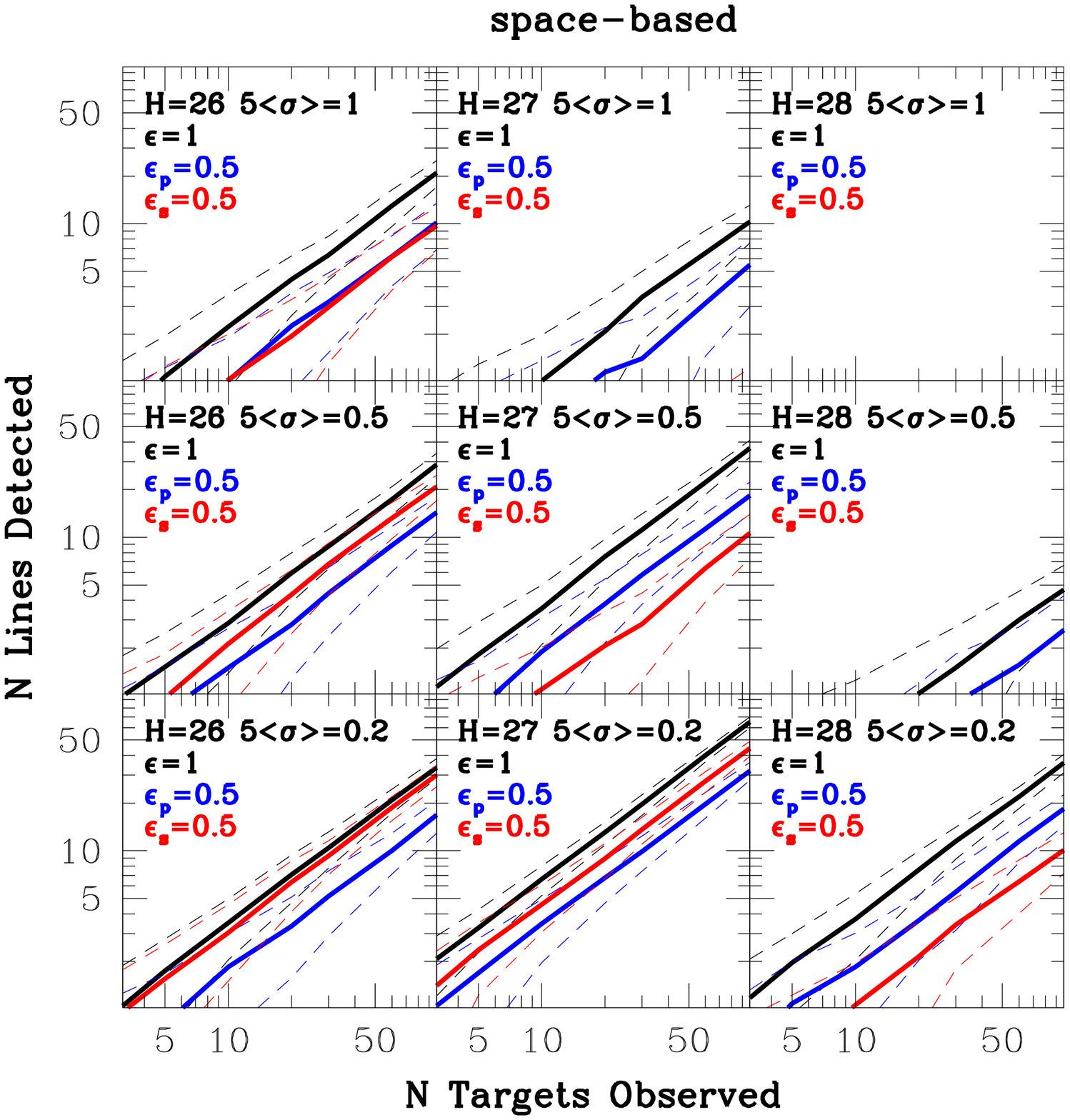}}
\end{center}
\figcaption{As in Figure~\ref{fig:ground} for space based observations.
\label{fig:space}}
\end{figure}

Figures~\ref{fig:ground} and~\ref{fig:space} show the number of
detections expected as a function of observed targets, together with
the r.m.s. scatter, as measured from Montecarlo simulations. The noise
level of the space based observations is uniform and equal to the
median value of the ground based observations, and they are given in
terms of 5$<\sigma>$ flux limits in units of 10$^{-17}$
ergs$^{-1}$cm$^{-2}$ in the captions. The brighter one is comparable
to that achieved by our 2.5hrs-long Keck-NIRSPEC integrations while
the fainter one is 5 times more sensitive. The fainter limits can be
achieved with realistically long observations with high efficiency IR
spectrographs on ground based 8-10m telescopes \citep[e.g.][reached
$5\cdot10^{-18}$ergs$^{-1}$cm$^{-2}$ in 14.8hrs of integration with
SINFONI]{Leh++10}, or with long multi-orbit integrations using the
grism mode on
WFC3-IR~\citep[see,e.g.,][]{Ate++11,Tru++11}. Alternatively, these
limits can be readily reached with the aid of moderate lensing
magnification, which is commonly found in the field of rich clusters
\citep[e.g.][]{Bra++09,Hal++11,Bra++08,Ric++08,Bra++11}. The detection rate is 
somewhat higher in general from space, since the sky emission lines
cause higher incompleteness in ground based data. A disadvantage of
WFC3 grism data is their low spectral resolution, compared to what is
generally obtained from the ground, and therefore the inability to
infer and use line shape information.

The average number of detected objects is a strong function of both
sensitivity and continuum magnitude. By going deeper one targets
intrinsically fainter objects, therefore with a higher fraction of
emitters, at the price of a higher noise in terms of $W$. In addition
the predictions of the patchy and smooth model differ significantly as
a function of depth and sensitivity. This is illustrated very clearly
in the middle row of Figures~\ref{fig:ground}
and~\ref{fig:space}. With 5-$\sigma$ sensitivity of
$5\cdot10^{-18}$~erg s$^{-1}$cm$^{-2}$ at $H=26$ the smooth model
yields significantly more detections than the patchy
model. Conversely, at $H=27$, the patchy model yields more detections,
because the high equivalent width tail dominates at the fainter
magnitudes. At $H=28$, one has to go even deeper ($2\cdot10^{-18}$~erg
s$^{-1}$cm$^{-2}$) to have any realistic chances of detection.

The r.m.s. scatter in the predicted number of detections provides
additional insight into future strategies. First, it can be used to
estimate the minimum number of targets that one needs to observe to
have a detection. Depending on the model and depth of observations,
the minimum number of targets required for a detection (with $>84$
probability, i.e. from the 1-$\sigma$ lower limit) varies between a
few (for $\epsilon=1$, $H=27$ and depth $0.2$) and virtually infinity
for shallower observations at $H=28$. Second, it can be used to
estimate the minimum number of targets needed to distinguish between
models. At depths comparable to this present study, of order 60
targets are needed to distinguish $\epsilon=1$ from $\ep$ or
$\es=0.5$. In the more favorable case of deeper observations (0.5
depth) at $H=27$, of order 20 targets would be sufficient for that
purpose, while $\sim50$ or more would be needed to start to
distinguish between $\ep=0.5$ and $\es=0.5$.

Clearly, at the moment, we are far from having a number of detections
at $z\sim8$ sufficient to characterize the distribution of
Lyman$\alpha$ emission, and in turn the properties of galaxies and the
intergalactic medium at that time. However, this goal is within reach
in the next few years. To evaluate an efficient strategy we need to
consider the density of Y-band dropouts in the sky as a function of
magnitude. Those are highly uncertain at this time, especially at the
bright end of the luminosity function, therefore we can only consider
them as rough estimates. We consider two estimates of the differential
number count densities, based on the luminosity function
\citep[$\phi_*=0.38\cdot10^{-3}$Mpc$^{-3}$; $\alpha=-2.0$; M$_*=-20.3$][]{Bou++11b} and on the observed
counts of~\citet{Bou++11b} and on our own estimate from BoRG at the
bright end~\citep{BORG1}. The lower estimates come from observed
number counts and the higher estimates from the luminosity function,
i.e. corrected for incompleteness.  The resulting differential number
count densities are 0.04, 0.3, 1.2, 2.4 arcmin$^{-2}$ mag$^{-1}$ and
0.05, 0.4, 1.8, 5.4 arcmin$^{-2}$ mag$^{-1}$, respectively at
$H=26,27,28,29$. These densities correspond to roughly 0.18-0.23,
1.4-2.0, 5.6-8.4, 11-26 per WFC3 field of view, and 1.4-1.8, 11-14,
43-65, 86-194 per MOSFIRE field of view \citep{McL++10}. Thus, in
blank fields, WFC3 effectively does not provide any multiplexing
advantage until $H\sim28$, where hope of detection starts at
$5\cdot10^{-18}$~erg s$^{-1}$cm$^{-2}$. This requires deep $\sim$20
orbits-long integration according to the WFC3 exposure time
calculator. Conversely, it is sufficient to reach beyond $H=27$ to
start gaining significantly with MOSFIRE, neglecting the positive
effects of clustering \citep{BORG2}. Even with moderate gravitational
lensing magnification $\mu$ one gains substantially in
multiplexing. The gain is especially marked at the bright end, where
the number counts are dominated by the exponential part of the
luminosity function~\citep[e.g.][]{Tre10}, and therefore the
differential surface density increases as $e^{\mu}/\mu$, i.e. a factor
of $\sim5$ per magnitude. In addition, by effectively going deeper one
further gains from the higher fraction of Lyman$\alpha$ emitters
amongst the intrinsically fainter population of galaxies (see
Figure~\ref{fig:models}).  An accurate estimate of the lensing gain
will depend on the details of the gravitational telescope under
consideration and is beyond the scope of this paper. In the longer
run, the James Webb Space Telescope will be able to detect
significantly fainter emission. In eight hours of integration with the
G140M grism $10^{-18}$~erg s$^{-1}$cm$^{-2}$ can be detected at
S/N=9. JWST can even detect the continuum of these sources, if Lyman
$\alpha$ is completely absent. At AB magnitude $26$ within the G140M
grism bandpass, NIRSPEC can detect {\it the continuum} with S/N=3 per
resolution element in eight hours.

\section{Summary}
\label{sec:conc}

With the goal of understanding the properties of the first galaxies
and the intergalactic medium at $z\sim7$ and above, we have developed
a simple yet powerful Bayesian framework to analyze observations of
Lyman$\alpha$ in emission. The framework is flexible enough to enable
the combination of datasets of different completeness, with different
noise properties. In addition, it enables one to take full advantage
of the information available.

Within this framework we implement two simple phenomenological models
to describe the evolution of the distribution of equivalent widths
with respect to a reference distribution, the one measured at $z\sim6$
by Stark et al.\ (2011). In the patchy model, equivalent to that
considered by previous work \citep{Fon++10,Ono++11,Sch++11,Pen++11},
Lyman$\alpha$ at $z>6$ is either completely absent or drawn from the
$z\sim6$ distribution (with probability $\ep$). In the smooth model,
the distribution of Lyman$\alpha$ is homogeneously reduced by a factor
$\es$. These models can be thought as simple idealizations of patchy
and smooth reionization. In the first case, some of the line of sights
are completely absorbed by the intergalactic medium, while others are
unabsorbed. In the second case, every line of sight is attenuated by
the same amount. Clearly, reality is likely to be more complicated,
but these two models should bracket somewhat the expected behavior of
the IGM near the epoch of reionization and therefore provide useful
guidance in planning observations and interpreting data.  The
parameters $\ep$ and $\es$ can be physically interpreted as the
average excess optical depth of Lyman$\alpha$ with respect to
$z\sim6$, i.e. $\langle e^{-\tau_{Ly\alpha}}\rangle$.

We apply our methodology to a sample of 39 $z\sim7$ dropouts collected
from the literature and to new and published observations of $z\sim8$
dropouts. Our findings can be summarized as follows:

\begin{itemize}
\item At $z\sim7$ the distribution of Lyman$\alpha$ equivalent 
width is significantly reduced with respect to $z\sim6$, consistently
for the patchy and smooth model, respectively by factors
$\es=0.69\pm0.12$ and $\ep=0.66\pm0.16$. The data do not provide
enough information to choose between our two models.
\item The models can be used to compute fractions of emitters above any
equivalent width $W$ at $z\sim7$. For $W>25$\AA, we find
$X^{25}_{z=7}=0.37\pm0.11$ ($0.14\pm0.06$) for galaxies fainter
(brighter) than M$_{\rm UV}$=-20.25 for the patchy model. This is
consistent with previous work, but with a smaller uncertainties by
virtue of our full use of the data. For the smooth model we find
$X^{25}_{z=7}=0.14\pm0.06$ and $X^{25}_{z=7}=0.38\pm0.11$,
respectively for the bright and faint subsample.

\item We observed with the Keck Telescope a bright and spatially 
resolved Y-band dropout (H$\approx26$), selected as part of the BoRG
survey \citep{BORG1}. We do not detect any emission lines down to a
5$-\sigma$ limit of 10$^{-17}$erg s$^{-1}$cm$^{-2}$. The lack of
emission lines eliminates the possibility that this galaxy is a pure
emission line object at lower redshifts.

\item At $z\sim8$ we combine our new observations with those of three 
dropouts observed by \citet[][including one target from our own BoRG
Survey]{Sch++11} and by \citet{Leh++10} and find that the inference is
inconclusive.

\item We forecast the outcome of  future observations of $z\sim8$ galaxies as a function of continuum magnitude and spectroscopic sensitivity, and show that it is possible to detect Lyman$\alpha$ and start to constrain its distribution by observing several tens of targets.

\end{itemize}

In conclusion -- even though much progress has been made at $z\sim7$
and on the imaging front at $z\sim8$ -- more spectroscopic data are
clearly needed to characterize the elusive population of $z\sim8$
galaxies and the distribution of Lyman $\alpha$ emission and
absorption. Our models show that making progress will require
substantial effort, even with sensitivities within reach of the grism
mode on board WFC3 and upcoming infrared spectrographs such as
MOSFIRE.  However, progress is definitely within reach, especially
with the assistance of lensing magnification provided by clusters of
galaxies used as gravitational telescopes.

\acknowledgments

Some of the data presented herein were obtained at the W.M. Keck
Observatory, which is operated as a scientific partnership among the
California Institute of Technology, the University of California and
the National Aeronautics and Space Administration. The Observatory was
made possible by the generous financial support of the W.M. Keck
Foundation. The authors wish to recognize and acknowledge the very
significant cultural role and reverence that the summit of Mauna Kea
has always had within the indigenous Hawaiian community.  We are most
fortunate to have the opportunity to conduct observations from this
mountain.  This paper is also based on observations made with the
NASA/ESA Hubble Space Telescope, obtained at the Space Telescope
Science Institute, which is operated by the Association of
Universities for Research in Astronomy, Inc., under NASA contract NAS
5-26555.  These observations are associated with program \#11700.
Support for program \#11700 was provided by NASA through a grant from
the Space Telescope Science Institute, which is operated by the
Association of Universities for Research in Astronomy, Inc., under
NASA contract NAS 5-26555. T.T. acknowledges support by the Packard
Foundation through a Packard Fellowship, and useful conversation with
B.~J.~Brewer and P.~J.~Marshall about Bayesian Statistics. T.T. thanks
A.~Pickles of assistance in using his spectral library and for
developing a most useful java applet. TT gratefully acknowledges the
hospitality of the Space Telescope Science institute funded by the
Distinguished Visitor Program, and of the Osservatorio Astronomico di
Roma, where parts of this paper were written. We thank Laura
Pentericci, Adriano Fontana, Marco Castellano, Maru{\v s}a Brada{\v
c}, and Kristian Finlator for useful suggestions. We thank the referee
for comments that improved the manuscript.

\appendix
\section{Alternative parameterization of $z\sim6$ W distribution}

We consider here an alternative parameterization of the distribution of
W at $z\sim6$ and derive the relevant formulae in this case. If, as
suggested by \citep{Fon++10,Pen++11}, the distribution of $W$ for
faint sources at $z\sim6$ has an additional tail of high equivalent
width distributions, represented by a uniform distribution out to
$W_m$=150\AA, Equation 1 becomes:

\begin{equation}
p_6(W)=\frac{2 A}{\sqrt{2 \pi}W_{c}}e^{-\frac{1}{2}\left(\frac{W}{W_{c}}\right)^2}H(W)+(1-A-B)\delta(W)+\frac{B}{W_m} H(W)H(W_m-W).
\end{equation}

By fitting the distribution measured by \citet{SEO11}, we find that
$A=0.81$, $B=0.05$ and $W_c=43$\AA\ provide a good description of the
data. $W_c$ is somewhat reduced with respect to the default case to
counterbalance the extra uniform tail. Then
Equations~\ref{eq:ppz}~and~\ref{eq:psz} gain an additional term of the
likelihood:

\be
p_{pu}(\ep,z_i|\{f,\sigma\},m)=p_p+\frac{C}{2ZW_m}\left(B\ep\left(\rm{erfc}(-\frac{f_i}{\sqrt{2}\sigma_i})-\rm{erfc}(\frac{W_m f_m - f_i}{\sqrt{2}\sigma_i})\right)\sqrt{2\pi}\sigma_i e^{\frac{1}{2}\left(\frac{f_i}{\sigma_i}\right)^2} \right) p(z_i),
\ee
\be
p_{su}(\es,z_i|\{f,\sigma\},m)=p_s+\frac{C}{2ZW_m}\left(B\left(\rm{erfc}(-\frac{f_i}{\sqrt{2}\sigma_i})-\rm{erfc}(\frac{\es W_m f_m - f_i}{\sqrt{2}\sigma_i})\right)\sqrt{2\pi}\sigma_i e^{\frac{1}{2}\left(\frac{f_i}{\sigma_i}\right)^2} \right) p(z_i),
\ee
where 1-A$\ep$ needs to be replaced with 1-(A+B)$\ep$ and 1-A needs to be replaced with 1-A-B in $p_p$ and $p_s$.

Similarly for the flux data case,
Equations~\ref{eq:ppW}~and~\ref{eq:psW} become:

\be
p_{pu}(\ep|W_o)=p_p+\frac{B\ep}{2ZW_m}\left(\rm{erfc}(-\frac{W_o}{\sqrt{2}\sigma_W})-\rm{erfc}(\frac{W_m-W_o}{\sqrt{2}\sigma_W})\right)p(\ep),
\ee
\be
p_{su}(\es|W_o)=p_s+\frac{B}{2ZW_m}\left(\rm{erfc}(-\frac{W_o}{\sqrt{2}\sigma_W})-\rm{erfc}(\frac{\es W_m-W_o}{\sqrt{2}\sigma_W})\right)p(\es)
\ee
where again 1-A$\ep$ needs to be replaced with 1-(A+B)$\ep$ and 1-A
needs to be replaced with 1-A-B in $p_p$ and $p_s$.

\begin{figure}
\begin{center}
\resizebox{0.97\columnwidth}{!}{\includegraphics{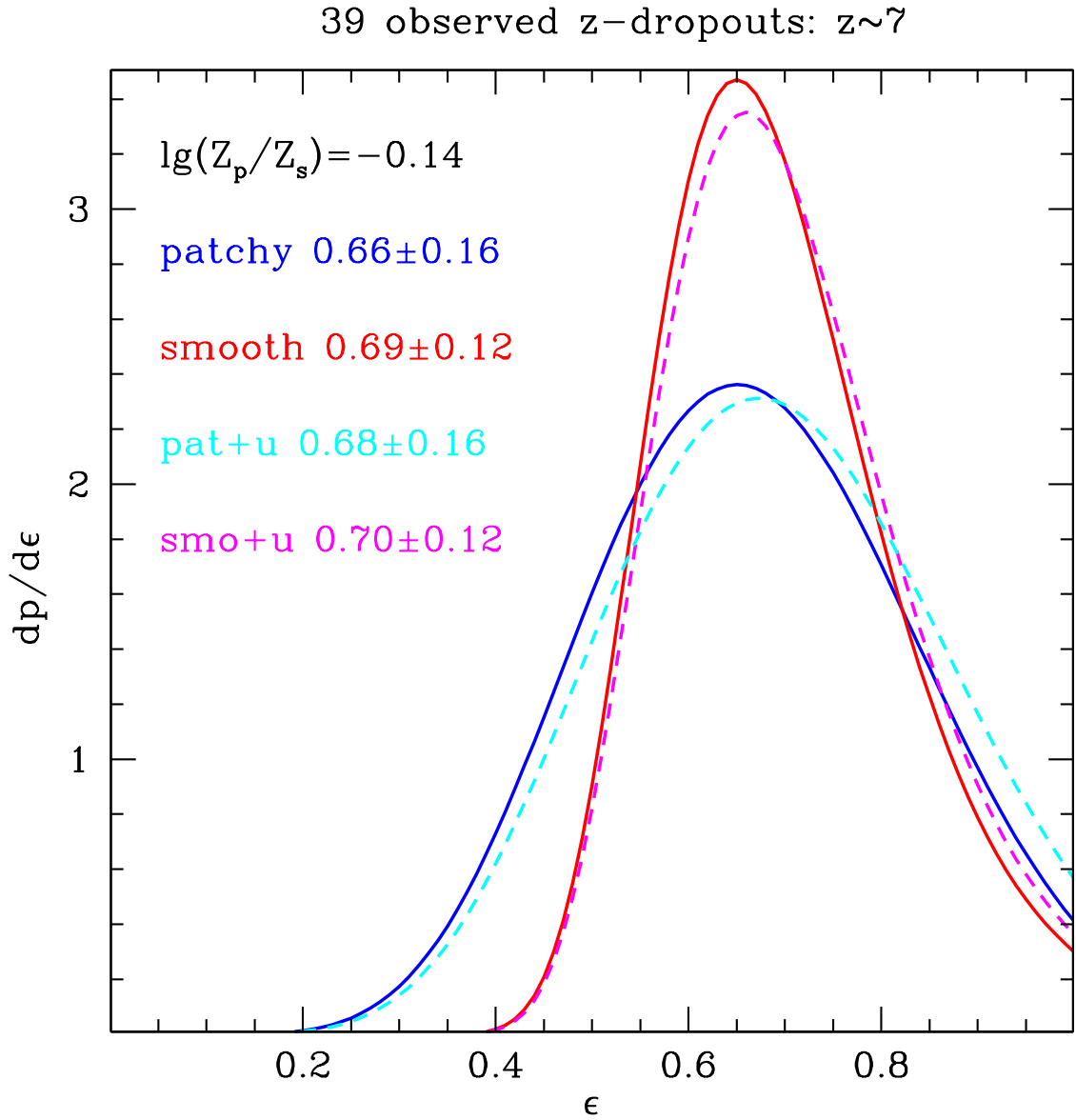}}
\end{center}
\figcaption{As Figure~\ref{fig:zdrops}, with the addition of two models
that include a uniform tail extending to W=150\AA\ for faint galaxies
at $z\sim6$ (magenta and cyan dashed lines). Our conclusions are
unchanged.
\label{fig:z7u}}
\end{figure}

As a test, we repeat the inference on $z\sim7$ galaxies using this
modified distribution of W for faint galaxies at $z\sim6$. The results
are shown in Figure~\ref{fig:z7u}~are well within the errors of the
inference with our default choice. The evidence ratio does not express
a preference for the default choice or the one with the extra uniform
tail (evidence ratio $<0.12$ dex between).

\bibliographystyle{apj}

\end{document}